\documentclass{article}

\usepackage{arxiv}

\usepackage[utf8]{inputenc} 
\usepackage[T1]{fontenc}    
\usepackage{hyperref}       
\usepackage{url}            
\usepackage{booktabs}       
\usepackage{amsfonts, amsmath}       
\usepackage{nicefrac}       
\usepackage{microtype}      
\usepackage{lipsum}		

\usepackage{float}
\usepackage{pdflscape}

\usepackage{graphicx}
\usepackage[square,comma,numbers]{natbib} 
\usepackage{doi}
\graphicspath{{figures/}}
\usepackage{subfig}

\title{Nombre Effectif de Partis Politiques en Afrique : Une Nouvelle Méthode pour un Calcul Objectif et Institutionnellement Neutre}

\author{ \href{https://orcid.org/0000-0003-0324-1443}{\includegraphics[scale=0.06]{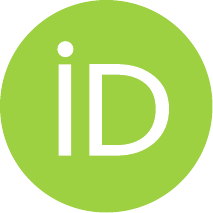}\hspace{1mm} Adama~Nouboukpo}
        \thanks{Corresponding author: Adama Nouboukpo, PhD Email: \texttt{noua06@uqo.ca, edu.nouboukpoadama@gmail.com}.} \\
        Département d'informatique et d'ingénierie  \\
        Université du Québec en Outaouais \\
        283 Boul. Alexandre-Taché
        Gatineau, QC J8X 3X7, Canada  \\
        \texttt{noua06@uqo.ca} \\
        \And
	\href{https://orcid.org/0009-0009-0755-6523}{\includegraphics[scale=0.06]{orcid.pdf}\hspace{1mm} Kodzo Michel Aladji} \\
        Département d'informatique et d'ingénierie  \\
        Université du Québec en Outaouais \\
        283 Boul. Alexandre-Taché
        Gatineau, QC J8X 3X7, Canada  \\
        \And
        \href{https://orcid.org/0009-0002-7235-8653}{\includegraphics[scale=0.06]{orcid.pdf}\hspace{1mm} Bappa Muktar} \\
        Département d'informatique et d'ingénierie  \\
        Université du Québec en Outaouais \\
        283 Boul. Alexandre-Taché
        Gatineau, QC J8X 3X7, Canada  \\
}

\renewcommand{\shorttitle}{Vers un Indice de Neutralité Institutionnelle}

\hypersetup{
pdftitle={A template for the arxiv style},
pdfsubject={q-bio.NC, q-bio.QM},
pdfauthor={David S.~Hippocampus, Elias D.~Striatum},
pdfkeywords={First keyword, Second keyword, More},
}

\begin{document}
\maketitle

\begin{abstract}
La fragmentation politique en Afrique constitue un défi important pour la gouvernance efficace et la stabilité. Les mesures traditionnelles de la fragmentation du système de partis, telles que l'indice du nombre effectif de partis (NEP), ne parviennent souvent pas à saisir les réalités nuancées des paysages politiques africains, en particulier l'influence des partis dominants, la fluidité des affiliations partisanes et l'impact des clivages ethniques et régionaux. Pour surmonter ces limitations, cet article introduit deux nouvelles mesures « apolitiques » ou «institutionnellement neutre» pour calculer le nombre effectif de partis, en se concentrant sur les dimensions géographiques et démographiques, notamment la taille de la population et la superficie du territoire. En intégrant ces réalités locales et en assurant un seuil minimum de deux partis, les modèles proposés offrent un cadre plus simple et plus pertinent sur le plan contextuel pour comprendre la dynamique politique en Afrique, en particulier dans les environnements où les données sont rares. Cette approche fournit un outil précieux pour analyser et rationaliser les systèmes politiques, avec un potentiel d'application plus large au-delà du contexte africain.
\end{abstract}

\keywords{Fragmentation politique \and Nombre effectif de partis \and Indice apolitique \and Systèmes politiques africains \and Dynamique des partis \and Données rares }

\section{Introduction} 
La fragmentation politique  se caractérise par une multiplication des partis et des groupes d'opinion au sein du paysage politique, rendant difficile la formation de majorités et compromettant ainsi l'efficacité de la gouvernance \cite{ricciuti2004, pildes2021p}. En Afrique, cette fragmentation revêt un caractère complexe et multidimensionnel, marqué par la division des États en multiples entités politiques souvent alignées sur des clivages ethniques, religieux ou régionaux. 
Cependant, si la fragmentation politique demeure modérée dans les pays anglophones et lusophones, elle s’avère nettement plus marquée dans l’espace francophone, où une multitude de partis nourrit l’ambition de conquérir le pouvoir. 
Par exemple, la République démocratique du Congo (RDC) détient le record parmi les pays africains en termes de nombre de partis politiques, avec un total de 910 selon la liste des partis autorisés \cite{RDCCENI} en vigueur en RDC en juin 2023. Toutefois, certains de ces partis, bien qu’officiellement enregistrés, n’ont pas de présence ou d’influence significative au niveau national. Si la multiplicité des partis est souvent considérée comme un signe de démocratie, elle conduit à une dispersion des suffrages et rend difficile la formation de majorités cohérentes \cite{koffielections, tine2017senegal, stoner2013t}. Les partis politiques africains sont généralement fragiles, dotés de structures et de programmes faibles, ne représentant pas des mouvements de masse et souffrant d’un manque d’organisation stable ainsi que d’activistes engagés \cite{erdmann2008party}. De plus, nombreux sont les partis qui manquent de programmes clairs et se concentrent souvent autour de personnalités plutôt que sur des idéologies ou des politiques publiques \cite{Osei2006}. Bratton et van de Walle \cite{bratton1997} soulignent que de nombreux partis en Afrique sont centrés sur des personnalités plutôt que sur des idéologies, ce qui limite leur capacité à représenter efficacement les citoyens. Cette situation affaiblit la capacité des États à répondre aux besoins de la population et à mettre en œuvre des réformes structurelles. Par conséquent, cette fragmentation excessive des partis politiques en Afrique constitue un obstacle à une gouvernance efficace et à la stabilité politique.\\ 

Un nombre limité et raisonnable de partis conduit à un système hautement institutionnalisé, favorisant la stabilité politique et le développement démocratique \cite{Daddieh2014}. L’institutionnalisation implique des structures organisationnelles solides, des idéologies claires et une présence constante sur la scène politique, contribuant ainsi à la stabilité du processus démocratique. Une approche triviale pour réduire la fragmentation des systèmes politiques consisterait à encourager leur regroupement par régions ou par zones géographiques qu’ils représentent, ou selon leurs idéologies \cite{collier2009}. Dans les pays caractérisés par de fortes divisions régionales, les partis pourraient former des coalitions régionales plutôt que de maintenir des structures séparées, ce qui permettrait de réduire le nombre de partis tout en assurant une représentation régionale équilibrée. Par ailleurs, l’exigence d’une présence significative dans plusieurs régions pour obtenir une reconnaissance nationale empêcherait la création de partis dits \emph{"Boutique"}, ne représentant que des intérêts locaux ou personnels \cite{bayart2009}. Cependant, réduire le nombre de partis nécessite un cadre méthodologique robuste pour quantifier la fragmentation politique.\\


L’indice du \emph{nombre effectif de partis} (NEP), introduit par Laakso et Taagepera \cite{Laakso1979}, constitue une pierre angulaire en science politique pour mesurer la fragmentation des systèmes partisans. 
Cet indice  mesure l’influence réelle des partis dans un pays, en tenant compte de la répartition des sièges ou des voix obtenues par chacun des partis politiques lors d’une élection. Toutefois, il présente plusieurs limitations, telles qu’une fragmentation maximale exagérée, l’incapacité à distinguer une domination absolue, une sensibilité aux petits partis, un manque de considération idéologique et surtout est biaisé si l'on utilise des données erronnées. Par ailleurs, si les données sur les élections sont inexistances, on ne peut pas calculer cet indice. Ces insuffisances ont conduit à diverses alternatives et modifications de la formule originale du NEP. De nombreux travaux \cite{taagepera1985rethinking, taagepera1993predicting, blau2008effective, caulier2005effective, grofman2012many, bhattacharya2006effective} dans ce domaine se sont concentrés sur le raffinement du concept d’indice du NEP et de ses méthodes de calcul. Ils ont souligné la nécessité de prendre en compte des facteurs allant au-delà des voix et des sièges, tels que le pouvoir législatif, les positions au sein du gouvernement et les similarités idéologiques entre partis. Certains chercheurs ont proposé de nouvelles interprétations du NEP, comme le cadre statistique de Caulier \cite{caulier2005effective} qui le lie à la taille des partis par un échantillonnage biaisé par la taille. D’autres, tels que Grofman et Kline \cite{grofman2012many}, ont introduit des mesures permettant de tenir compte des regroupements idéologiques. Parmi les contributions notables visant à améliorer l’indice NEP, on peut citer la nouvelle approche de Golosov \cite{Golosov2010} pour les systèmes fragmentés et concentrés, les indices d’inégalité d’entropie généralisée de Borooah \cite{borooah2013general} et le \emph{Seat Product Model} de Taagepera \cite{taagepera2007}. Malgré ces avancées, les débats persistent quant à la meilleure manière de représenter la pertinence des partis et la fragmentation des systèmes, notamment dans les situations de majorité à parti unique et dans les systèmes fortement fragmentés.\\

Une analyse centrée uniquement sur les systèmes électoraux peut négliger d'autres facteurs déterminants, tels que les clivages sociaux ou la culture politique. De plus, les mesures conçues pour des contextes de majorité à parti unique sont souvent moins adaptées aux systèmes caractérisés par une fragmentation des partis \cite{raymond2016e}. Plus encore, aucun indice unique ne peut capturer pleinement la complexité des systèmes de partis, car la mesure la plus appropriée dépendant souvent des objectifs spécifiques de recherche, de l’accessibilité des données et des facteurs contextuels \cite{raymond2016e,wolff2005e}. Ainsi, les chercheurs doivent déterminer si leur analyse porte sur le pouvoir de négociation, la répartition des suffrages ou la diversité idéologique \cite{wolff2005e}. En outre, la disponibilité et la fiabilité des données concernant le comportement des coalitions, l’influence législative ou les regroupements idéologiques jouent un rôle crucial dans le choix de l’indice le plus pertinent \cite{raymond2016e, hazama2003s}. Compte tenu de ces complexités, les méthodes récentes recourent fréquemment à plusieurs variantes de l’indice NEP — y compris celles basées sur les voix, les sièges et le pouvoir législatif — afin d’obtenir une compréhension plus complète des systèmes partisans \cite{raymond2016e}. Par ailleurs, le paysage politique, tel que le degré de fragmentation, la maturité des institutions démocratiques et la structure gouvernementale, influence également le choix de l’outil de mesure \cite{wolff2005e}.\\

Dans de nombreux pays africains, les modèles NEP conventionnels peinent à saisir les réalités nuancées des systèmes de partis \cite{bogaards2004}. Les mesures traditionnelles, telles que l’indice Laakso Taagepera, reposent fortement sur des parts de voix ou de sièges stables, alors que de nombreux environnements électoraux africains se caractérisent par une affiliation fluide, des alliances changeantes et même des irrégularités électorales. Dans ces contextes, les dynamiques politiques impliquent souvent des partis dominants qui, bien que peu nombreux, exercent un contrôle disproportionné via des réseaux de patronage et des pratiques clientélistes, des aspects que ces indices ne parviennent pas à capturer \cite{erdmann2008party, koffielections, bayart2009}. 
De plus, les systèmes électoraux en Afrique peuvent être hybrides ou en évolution, conduisant à des résultats volatils que les modèles statiques ne peuvent représenter avec précision.  Cette complexité est encore accentuée par la diversité ethnique et les disparités régionales, que les modèles NEP existant \cite{caulier2005effective} ne parviennent pas à pleinement intégrer. Or, ces clivages ethnorégionaux jouent un rôle déterminant dans la fragmentation politique du continent, limitant ainsi la portée des analyses traditionnelles. Par ailleurs, la diversité des systèmes électoraux et des cadres réglementaires relatifs aux partis politiques peut engendrer des dynamiques spécifiques aux systèmes des partis qui ne sont pas prises en compte par des modèles conçus pour d’autres contextes. Par exemple, depuis mai 2020, le Bénin a réduit son nombre de partis politiques, passant de plus d’une centaine à treize (13), une situation que les modèles existants ne parviennent pas à saisir. En outre, des modèles plus avancés intégrant le pouvoir législatif, les dimensions idéologiques ou des mesures basées sur l’entropie se heurtent à d’importants défis dans le contexte africain. Ces approches nécessitent généralement des données détaillées et fiables sur le cadre législatif, les accords de coalition ou la répartition des électeurs — des données \cite{blau2008effective} qui sont souvent rares ou incohérentes dans de nombreux pays africains. De plus, les hypothèses sous-jacentes de stabilité et de compétition rationnelle dans ces modèles ne tiennent souvent pas dans des contextes où les identités et les alliances politiques sont très dynamiques et influencées par des facteurs non institutionnels. 

En conséquence, les méthodes existantes échouent souvent à refléter la véritable complexité des systèmes de partis en Afrique, d’où la nécessité de développer de nouveaux outils de mesure adaptés sensibles au contexte \cite{bogaards2004}. Enfin, ces limitations soulignent le besoin de mesures du NEP plus intégrées et adaptables, capables de saisir la complexité des systèmes de partis politiques à travers divers contextes, en particulier en Afrique.\\


Pour remédier à ces limitations, nous proposons dans cet article deux formules innovantes et alternatives pour calculer le nombre effectif de partis, mieux adaptées aux réalités des paysages politiques africains. Ces formules sont \emph{apolitiques}, c'est-à-dire qu'elles ne dépendent ni du pouvoir de négociation, ni des répartitions de voix ou de sièges, ni de la diversité idéologique. Au lieu de cela, nos modèles se concentrent exclusivement sur des dimensions géographiques et démographiques, en intégrant des réalités locales telles que la taille de la population et la superficie du territoire afin de simplifier l’analyse des paysages politiques. Conçus pour être à la fois simples et efficaces, ces modèles intègrent également un seuil minimal de deux partis, reflétant la réalité politique universelle, même dans les environnements les moins compétitifs. Cette approche représente une avancée significative par rapport aux méthodologies existantes, offrant un cadre plus nuancé et contextuellement pertinent pour comprendre les dynamiques politiques. Nos contributions peuvent être résumées comme suit :

\begin{itemize}
\item Nous introduisons pour la première fois des modèles mathématiques \emph{"apolitiques"} permettant de calculer de manière significative le nombre effectif de partis, spécialement adaptés au contexte africain.

\item Contrairement aux approches se concentrant exclusivement sur les institutions ou les comportements électoraux, notre modèle intègre directement l’effet combiné de la démographie et de la géographie d’un pays.

\item Nos approches, simples, universelles et pragmatiques, reposent uniquement sur les réalités locales tout en garantissant un principe démocratique viable, ce qui en fait une solution précieuse en cas d’absence ou d’insuffisance de données institutionnelles.

\item Nous proposons un modèle flexible et adaptable, capable de refléter des contextes variés, tels que les grands pays à faible densité de population ou les petits pays à forte densité.

\item Enfin, bien que conçus pour le contexte africain, nos modèles sont facilement extensibles à d’autres régions du monde, offrant un outil universel pour analyser et rationaliser les systèmes politiques.
\end{itemize}

Le reste de cet article est organisé comme suit. La Section 2 passe en revue la littérature sur les métriques de calcul du nombre effectif de partis. La Section 3 présente nos modèles et le cadre d’estimation du nombre effectif de partis. La Section 4 détaille notre étude expérimentale, évalue la performance de nos méthodes et expose les résultats de leur application aux pays africains. Enfin, la Section 5 conclut notre travail et propose des pistes pour des recherches futures.

\section{Revue de la littérature sur les modèles du nombre effectif de partis}

Le concept de nombre effectif de partis \cite{Laakso1979}, introduit par Laakso et Taagepera en 1979, vise à quantifier le nombre de partis compétitifs dans un système électoral en tenant compte de leur poids relatif, c'est-à-dire en se basant sur la proportion de sièges ou de voix obtenus par les partis politiques. La formule est la suivante :

\begin{equation}
N_{LT} = \frac{1}{\sum_{i=1}^{n} s_i^2}
\end{equation}
où $s_i$ représente la proportion de sièges ou de voix obtenus par le $i$-ème parti et $n$ est le nombre total de partis. \\

Malgré son adoption généralisée, des critiques ont rapidement émergé concernant ses hypothèses sous-jacentes et son applicabilité à divers contextes — particulièrement dans les systèmes de partis très fragmentés ou très concentrés. Plusieurs approches ont cherché à généraliser ou formaliser le NEP à l’aide de nouvelles équations et de modèles basés sur l’entropie. Borooah et al. \cite{bogaards2004, borooah2013general} ont introduit une mesure générale utilisant des indices d’inégalité d’entropie généralisée, englobant ainsi les mesures existantes (y compris celle de Laakso–Taagepera) et permettant une agrégation au niveau national. Cette approche souligne comment des considérations subjectives (par exemple, la dimension d’« efficacité » que l’on privilégie) peuvent modifier substantiellement les valeurs du NEP. De même, Bhattacharya et Smarandache \cite{bhattacharya2006effective} ont développé un cadre d’équilibre politique entropique pour les démocraties multipartites avec « électeurs flottants », suggérant que le niveau d’incertitude ou de dispersion parmi les électeurs peut être capturé par un calcul du NEP basé sur l’entropie. D’un autre côté, Golosov \cite{Golosov2010} a proposé une nouvelle approche pour le calcul du NEP qui corrige les biais dans les contextes fortement fragmentés ou concentrés, améliorant ainsi la performance de l’indice dans différents environnements électoraux. Cet indice attribue un poids réduit aux petits partis afin de limiter leur impact sur le calcul du nombre effectif de partis. Sa formule est la suivante :

\begin{equation}
N_P = \sum_{i=1}^{n} \frac{1}{1 + \left( \frac{s_1^2}{s_i} \right) - s_i}
\end{equation}
où $s_i$ représente la proportion de sièges ou de voix obtenus par le $i$-ème parti et $s_1$ est la part du plus grand parti.

Bien que cette méthode soit efficace pour décrire des systèmes multipartites complexes, elle se concentre principalement sur des données « ex post » (résultats électoraux), ce qui limite son utilité pour prédire les effets des réformes institutionnelles ou pour des systèmes politiques émergents, en particulier en Afrique.

Les méthodes existantes qui se focalisent uniquement sur les voix ou les sièges bruts ne parviennent pas à capturer la véritable influence politique \cite{blau2008effective}. Caulier et Dumont \cite{caulier2005effective} prolongent l’idée des partis pertinents en intégrant des mesures du pouvoir de vote dans le NEP. Leur travail fournit un décompte plus nuancé qui reflète mieux la capacité réelle d’un parti à influencer les résultats législatifs ou exécutifs. De plus, Blau et al. \cite{blau2008effective} expliquent comment le NEP peut être mesuré à quatre échelles — voix, sièges, pouvoir législatif et pouvoir ministériel — en soulignant que de fortes parts de sièges ne se traduisent pas toujours par une influence politique proportionnelle. Puisque le NEP est étroitement lié au nombre de partis qui « comptent réellement », d’autres travaux examinent également la fragmentation des systèmes partisans et leur nationalisation. Kselman et Powell \cite{kselman2016crowded} étudient comment la fragmentation des partis peut faciliter l’entrée de nouveaux partis dans les démocraties établies, reliant ces dynamiques aux valeurs observées du NEP. Golosov \cite{golosov2015} replace le NEP dans le contexte plus large de la nationalisation des systèmes, montrant comment l’hétérogénéité territoriale peut influencer la fragmentation et la répartition du soutien aux partis. Xhaferaj \cite{xhaferaj2014} applique le concept du NEP au système politique albanais, soulignant l’importance de mesures spécifiques au contexte pour les démocraties naissantes. Grofman et Kline \cite{grofman2012many} proposent une mesure du nombre de partis idéologiquement reconnaissables, reflétant le degré selon lequel les partis peuvent être regroupés sur la base de positions politiques ou idéologiques cohérentes.\\

Récemment, le « Seat Product Model » \cite{taagepera2007}, un modèle mathématique fondé sur des principes déductifs, a proposé des modèles prédictifs et théoriques reliant le NEP aux propriétés plus larges du système électoral, telles que la magnitude des circonscriptions, les effets mécaniques des règles électorales et l’hétérogénéité de l’électorat. Ce modèle prédit le nombre effectif de partis dans une assemblée en se basant sur la magnitude des circonscriptions ($M$) et la taille de l’assemblée ($S$) :

\begin{equation}
N_s = (MS)^{1/6}
\end{equation}

Cette formule repose sur l’idée que la fragmentation politique dépend de la permissivité du système électoral. Elle se distingue par sa simplicité et son indépendance vis-à-vis des variables sociales ou comportementales, ce qui en fait un outil universellement applicable pour la conception institutionnelle. Cependant, ses limites incluent l’exclusion de systèmes complexes (multiniveaux) et l’incapacité à prédire le nombre de partis obtenant des voix ($N_s$). Li et Shugart \cite{li2016seat} étendent le modèle du Seat Product pour inclure des systèmes électoraux multiniveaux (avec des sièges compensatoires) et intègrent des données issues d’élections dans des démocraties établies et émergentes. Leur modèle conserve la structure institutionnelle simple de Taagepera tout en ajoutant un facteur correctif pour les systèmes complexes, avec la formule suivante :

\begin{equation}
N_s = K^t (MS_B)^{1/6}
\end{equation}

où $K_t$ est une constante déterminée empiriquement, liée à la proportion de sièges attribués au niveau supérieur ($t$), et $MS_B$ représente le produit de la magnitude des circonscriptions et de la taille totale des sièges de base. Ils démontrent également que l’ajout de variables sociales, telles que la diversité ethnique, n’améliore que marginalement la précision des prévisions. \\

En résumé, trois approches majeures ont été largement utilisées, à savoir l’indice Laakso–Taagepera \cite{Laakso1979}, le modèle de Golosov \cite{Golosov2010} et le Seat Product Model de Taagepera \cite{taagepera2007}. Ces trois modèles présentent des approches complémentaires pour mesurer et prédire le nombre effectif de partis. L’indice Laakso–Taagepera demeure une référence pour sa simplicité, tandis que Golosov apporte une correction utile pour les systèmes très concentrés. Enfin, le modèle « Seat Product » de Taagepera offre une perspective institutionnelle robuste. Toutefois, des lacunes subsistent, notamment en ce qui concerne la prise en compte des dynamiques comportementales et des clivages sociaux. L’intégration des forces et des limites de ces modèles pourrait ouvrir la voie à des outils plus complets pour analyser les systèmes politiques.

\section{Méthodes proposées} 
Notre approche introduit une nouvelle méthode de calcul du nombre effectif de partis basée sur deux facteurs essentiels : la taille de la population et la superficie des pays ou des régions. Cette méthode, que nous appelons \textbf{les indices de Noua}, vise à fournir une compréhension plus globale des systèmes de partis politiques en intégrant des variables démographiques et géographiques. Nous proposons deux modèles à savoir le modèle « hard » et le modèle « soft ». Ces deux modèles garantissent un minimum de deux partis politiques (\( NEP \geq 2 \)), reflétant la réalité politique universelle selon laquelle même les systèmes les moins compétitifs nécessitent une forme d’opposition. \\

Soit \( P \) la population et \( A \) la superficie d’un pays donné. Nous introduisons des constantes ajustables \( k_1 \) et \( k_2 \) pour pondérer l’importance relative de la population et de la superficie dans la détermination du nombre de partis. Plus précisément, \( k_1 \) représente le nombre d’habitants par parti politique, tandis que \( k_2 \) représente la superficie par parti politique.

\subsection{Le modèle hard}
Le modèle hard calcule le nombre effectif de partis en utilisant une transformation logarithmique de la population et de la superficie, tout en garantissant un minimum de deux partis. Le modèle hard est défini par l’équation suivante :

\begin{equation}
\text{NEP} = \max\left(2, \log\left(\frac{P}{k_1}\right) + \log\left(\frac{A}{k_2}\right)\right)
\end{equation}

où \(\log\) désigne le logarithme naturel.\\

Pour ce modèle, comme la population est exprimée en millions et la superficie en milliers, nous utilisons \( k_1 = 1\,000\,000 \) et \( k_2 = 1000 \). Lorsque \( k_1 \neq 1\,000\,000 \) et \( k_2 \neq 1000 \), nous désignons cette variante sous le nom de modèle Hard-Flex, adapté aux pays non africains.

\subsection{Le modèle soft}
Le modèle soft, quant à lui, utilise une combinaison linéaire de la population et de la superficie pour calculer le nombre effectif de partis. Le modèle soft est défini par l’équation :

\begin{equation}
\text{NEP} = \max\left(2, \frac{P}{k_1} + \frac{A}{k_2}\right)
\end{equation}

\hfill \break

Dans ce modèle, les paramètres \( k_1 \) et \( k_2 \) sont estimés à l’aide de l’algorithme de Levenberg-Marquardt (LM) \cite{ranganathan2004lm}, une technique d’optimisation largement utilisée pour résoudre des problèmes de moindres carrés non linéaires.  
Cet algorithme minimise l’erreur entre le nombre de partis prédit et le nombre de partis effectivement observé en ajustant itérativement les paramètres jusqu’à convergence.

\subsubsection{Estimation initiale des paramètres}

Nous avons initialisé l’algorithme d’optimisation avec des valeurs empiriques pour \( k_1 \) et \( k_2 \) dérivées des moyennes de la population, de la superficie et des valeurs du NEP obtenues par le modèle hard. Cette démarche repose sur le fait que \( k_1 \) et \( k_2 \) représentent respectivement le nombre d’habitants par parti politique et la superficie par parti politique. Cette estimation constitue un point de départ pertinent pour l’algorithme.

\begin{equation}
k_1^{(0)} = \frac{\text{mean}(P)}{\text{mean}(NEP^{observed})}, \quad
k_2^{(0)} = \frac{\text{mean}(A)}{\text{mean}(NEP^{observed})}
\end{equation}

où \( NEP^{observed} \) représente le nombre effectif de partis observé, dérivé du modèle hard. $\text{mean}(P)$ et $\text{mean}(A)$ caractérise  respectivement la moyenne de la population et de la superficie de tous les pays aricains.

\subsubsection{Processus d’optimisation}

L’algorithme minimise la somme des carrés des résidus entre les valeurs observées et prédites du NEP :

\begin{equation}
\min_{\{k_1, k_2\}} \sum_{i=1}^{n} \left[NEP_i^{observed} - \max\left(2, \frac{P_i}{k_1} + \frac{A_i}{k_2}\right)\right]^2
\end{equation}

où :
\begin{itemize}
    \item \( NEP_i^{observed} \) est la valeur du NEP observée pour le pays \( i \),
    \item \( P_i \) et \( A_i \) représentent respectivement la population en 2023 et la superficie du pays \( i \),
    \item \( n \) est le nombre total de pays officiellement constitués, qui est de 54 pour l’Afrique.\\
\end{itemize}

Le processus d’optimisation ajuste itérativement \( k_1 \) et \( k_2 \) jusqu’à ce que la convergence soit atteinte, définie par le fait que la variation de la somme des carrés des résidus entre deux itérations successives soit inférieure à un seuil prédéfini (1e-6 pour \( k_1 \) et 1e-4 pour \( k_2 \)). Afin d’assurer la robustesse de notre modèle, nous avons exécuté l’algorithme vingt (20) fois. Les valeurs optimales obtenues sont :

\begin{equation}
    k_1^{*} = 18\,819\,265; \quad
    k_2^{*} = 110\,014
\end{equation}

\subsection{Approche innovante de calcul du NEP avec les indices de Noua }

Nos modèles offrent une approche unique pour calculer le nombre effectif de partis en intégrant les facteurs de population et de superficie, lesquels ne sont pas habituellement pris en compte dans les calculs traditionnels du NEP \cite{li2016seat}. Alors que des méthodes établies, telles que l’indice Laakso–Taagepera, se concentrent sur les parts de voix ou de sièges, notre approche propose une perspective alternative basée sur des caractéristiques démographiques et géographiques. L’inclusion d’un seuil minimal de deux partis dans les deux modèles reflète la réalité pratique des systèmes politiques, où même les gouvernements fortement centralisés font généralement face à une forme d’opposition. Cette caractéristique garantit que nos modèles capturent l’essence de la compétition politique, même dans des systèmes politiques moins diversifiés.\\

En proposant à la fois un modèle hard et un modèle soft, notre approche permet une flexibilité d’application. Le modèle hard, avec son échelle logarithmique, peut être mieux adapté aux pays présentant des variations extrêmes de population ou de superficie, tandis que le modèle soft offre une relation plus linéaire entre ces facteurs et le NEP. L’utilisation de l’algorithme de Levenberg-Marquardt pour l’estimation des paramètres dans le modèle soft démontre notre engagement à optimiser la puissance prédictive du modèle. Cette approche basée sur les données renforce la fiabilité du modèle et son potentiel pour fournir des prévisions précises dans divers paysages politiques.

\section{Prédiction du nombre effectif de partis} 
Dans cette section, nous présentons la performance de nos modèles pour lesquels nous avons mené des expériences complètes afin de calculer le nombre effectif de partis. Les deux modèles ont été implémentés en Python à l’aide de \href{https://colab.google/}{Google Colab} \cite{gcolab}, fonctionnant sur un appareil équipé d’un processeur Intel Core i7-9750H @2.6GHz et d’un GPU NVIDIA. Nous utilisons l’algorithme \href{https://docs.scipy.org/doc/scipy/reference/generated/scipy.optimize.least_squares.html}{\emph{Levenberg-Marquardt}} de la bibliothèque SciPy \cite{scipy} pour déterminer les paramètres \( k_1 \) et \( k_2 \).

\subsection{Jeu de données}
\begin{landscape} 
\begin{figure}[H]
    \includegraphics[width=1\linewidth]{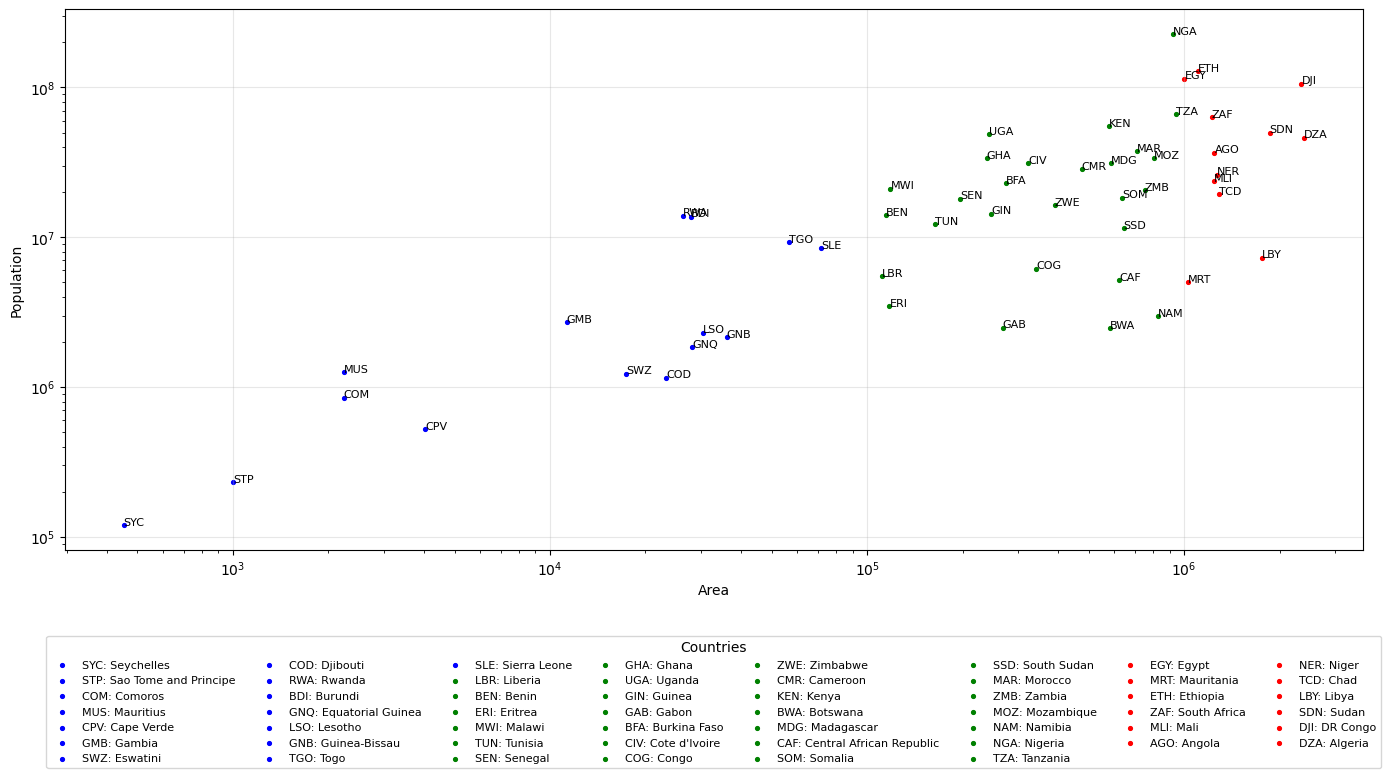}
    \caption{Population vs Superficie des pays africains}
    \label{fig:dataset}
\end{figure}
\end{landscape}

Cette section décrit les sources de données, les étapes de préparation et la structure finale du jeu de données utilisé dans cette étude. Les données relatives aux pays africains ont été extraites de la base de données du World Bank Group \cite{WorldBank2025}, en se concentrant uniquement sur les 54 nations africaines officiellement reconnues, incluant leurs statistiques de superficie et de population. La population est exprimée en millions d’habitants et la superficie en kilomètres carrés (km²). Le jeu de données a été constitué à partir de deux fichiers CSV contenant les données de population et de superficie pour les pays africains. Après l’importation des données, les deux sources ont été fusionnées en un seul jeu de données, garantissant ainsi la cohérence et ne conservant que les colonnes pertinentes : nom du pays, code du pays, population et superficie. Le code du pays a été utilisé pour distinguer, par exemple, les deux Congo et les trois Guinées. Nous avons sélectionné les données de population pour les années 2014 à 2023 et avons catégorisé chaque pays en fonction de sa région géographique : Afrique centrale, Afrique de l'Est, Afrique de l'Ouest, Afrique du Nord ou Afrique australe. De plus, nous avons inclus des communautés économiques régionales, telles que la Communauté économique des États de l’Afrique de l’Ouest (CEDEAO) \cite{ecowas}, la Communauté économique des États de l’Afrique centrale (CEEAC) \cite{eccas} et l’Alliance des États du Sahel (AES) \cite{allianceofsahelstates}. Le jeu de données final offre une base solide pour analyser la relation entre la population, la superficie et l’affiliation régionale des pays africains. La Table \ref{tab:dataset} présente un extrait du jeu de données final, tandis que la Figure \ref{fig:dataset} illustre la distribution des données.

\begin{table}[H]
\caption{Extrait du jeu de données final}
\label{tab:dataset}
\centering
\begin{tabular}{|c|c|c|c|c|}
\hline
Pays      & Population   & Superficie       & Région         & Communauté \\ \hline
RD Congo     & 109\,276\,265 & 2\,267\,050 & Afrique Centrale & CEEAC     \\ \hline
Togo        & 9\,304\,337  & 56\,785   & Afrique de l'Ouest   & CEDEAO    \\ \hline
Burkina Faso & 23\,548\,781  & 273\,600   & Afrique de l'Ouest    & AES       \\ \hline
Mali         & 24\,478\,595  & 1\,220\,190 & Afrique de l'Ouest    & AES       \\ \hline
Niger        & 27\,032\,412  & 1\,266\,700 & Afrique de l'Ouest   & AES       \\ \hline
\end{tabular}%
\end{table}

\subsection{Résultats des prédiction des indices de Noua}
\begin{landscape}
\begin{figure}[H]
    \includegraphics[scale=0.37]{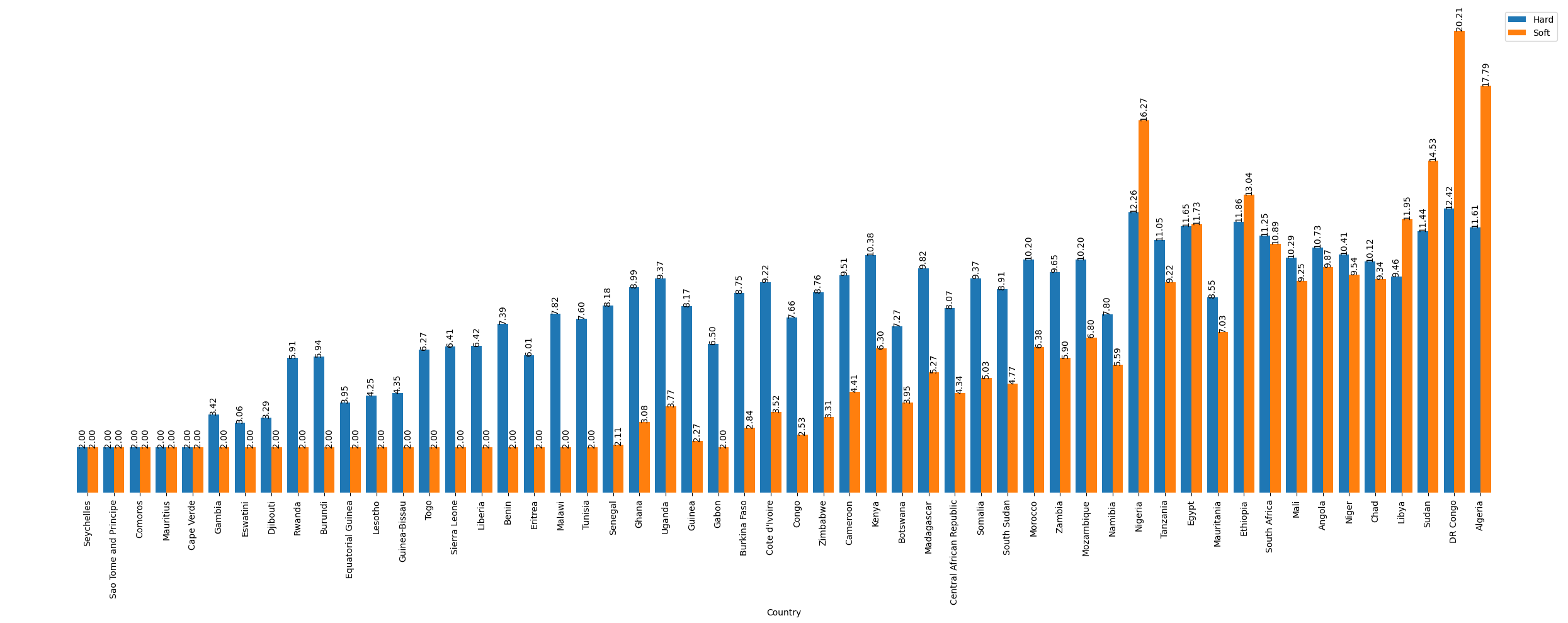}
    \caption{Comparaison du nombre effectif de partis entre les indices Hard et Soft à travers différents pays}   \label{fig:hard_vs_soft}
\end{figure}
\end{landscape}

Dans cette section, nous évaluons nos modèles \textit{Hard} et \textit{Soft} pour estimer le nombre effectif de partis politiques dans chaque pays d'Afrique. La Figure \ref{fig:hard_vs_soft} et la Table \ref{tab:NouaIndexes} présentent à la fois les tendances en matière de superficie/population et les différences entre nos deux approches. Dans la Table \ref{tab:NouaIndexes}, la dernière colonne représente la moyenne des valeurs obtenues avec les deux modèles afin d'avoir une vue d'ensemble du NEP pour un pays donné. Dans la Figure \ref{fig:hard_vs_soft}, les pays sont triés d'abord par superficie puis par population pour permettre une comparaison visuelle et mettre en évidence la distribution relationnelle de nos indices, ce qui facilite l'identification des relations entre les facteurs géographiques et démographiques ainsi que les variations spécifiques à chaque pays dans la relation entre les méthodes \textit{Hard} et \textit{Soft}. De manière générale, nous avons observé que la figure révèle des pointes notables dans chaque région, ces pointes se manifestant dans les plus grands pays en termes de superficie et de population pour les deux modèles. Par ailleurs, les pays de petite taille, tant en population qu'en superficie, affichent les valeurs du NEP les plus faibles.

\subsubsection{Observations sur le modèle Hard}

L'indice Hard du nombre effectif de partis, présenté dans la Figure \ref{fig:hard_vs_soft} (couleur Blue) et la Table \ref{tab:NouaIndexes} (deuxième colonne), révèle une variation significative entre les pays africains. Les petites nations (en termes de superficie), caractérisées par une faible population et une petite superficie, telles que les Seychelles, Sao Tomé-et-Principe, les Comores, Maurice et le Cap-Vert, affichent systématiquement la valeur minimale de 2,00. En revanche, les pays disposant d'une grande population et/ou d'une vaste superficie tendent à obtenir des scores plus élevés. Le Nigeria, avec la plus grande population et une superficie conséquente, illustre cette tendance avec une valeur élevée de 12,26. Cependant, la relation entre la population, la superficie et l'indice Hard n'est pas parfaitement linéaire. Bien que ces deux facteurs y contribuent, leur impact diminue de manière logarithmique, ce qui signifie que les augmentations de l'un ou l'autre variable entraînent des accroissements de l'indice de plus en plus faibles. Par exemple, alors que l'Algérie, la République démocratique du Congo et le Soudan disposent tous d'une très grande superficie, la population exceptionnellement élevée de la RDC lui permet d'atteindre le score le plus élevé (12,42). Aucun autre pays ne dépasse cette valeur, en raison de la combinaison en RDC de plus de 100 millions d'habitants et de plus de 2 millions de kilomètres carrés. Un groupe de pays, comprenant de grandes nations telles que le Soudan, l'Égypte, l'Afrique du Sud, l'Éthiopie et la Tanzanie, affiche des valeurs relativement élevées de l'indice Hard, généralement supérieures à 11. Cela suggère que, bien que ces pays puissent avoir des superficies supérieures à celle du Nigeria, leurs populations considérables tendent à réduire leur NEP. D'autres pays, notamment ceux d'Afrique centrale et d'Afrique de l'Ouest, présentent des valeurs intermédiaires, comprises entre 3 et 10, reflétant leur taille relative en termes de superficie et de population.

\subsubsection{Observations sur le modèle Soft}

L'indice Soft présente un schéma distinct par rapport à l'indice Hard. Un nombre significatif de pays se regroupe à l'extrémité inférieure de l'échelle, avec des valeurs d'indice Soft avoisinant 2,00. Ce groupe inclut les petites nations comme les Seychelles, São Tomé-et-Príncipe, les Comores, Maurice et le Cap-Vert, ainsi que des pays plus grands tels que la Gambie, Eswatini, Djibouti, le Rwanda, la Guinée équatoriale, le Lesotho, la Guinée-Bissau, le Togo, la Sierra Leone, le Libéria, le Bénin, l'Érythrée, le Malawi et la Tunisie. D'autres pays, présentant des valeurs relativement faibles de l'indice Soft (inférieures à 5,00), incluent le Sénégal, la Guinée, le Congo, le Burkina Faso, le Ghana, le Zimbabwe, la Côte d'Ivoire, l'Ouganda, le Botswana, la République centrafricaine, le Cameroun et le Soudan du Sud. Ce regroupement à l'extrémité inférieure suggère que l'indice Soft est moins sensible aux variations de superficie et de population. En revanche, les scores les plus élevés de l'indice Soft (supérieurs à 10,00) se trouvent principalement dans les pays plus grands et plus peuplés, tels que la République démocratique du Congo, l'Algérie, le Nigeria, le Soudan, l'Éthiopie, la Libye, l'Égypte et l'Afrique du Sud. Il est à noter que la République démocratique du Congo affiche la valeur la plus élevée, à 20,21. À l'exception de l'Afrique du Sud, ces valeurs dépassent celles de l'indice Hard. Les autres pays se situent dans une plage intermédiaire comprise entre 5,00 et 10,00.\\

Dans l'ensemble, l'indice Soft présente une corrélation plus faible avec les pays de petite et moyenne taille en termes de superficie et de population comparativement à l'indice Hard. Cependant, pour les pays plus grands, dotés d'une plus grande superficie et d'une population plus importante, l'indice Soft tend à produire des valeurs supérieures à celles de l'indice Hard. Cela souligne l'importance de prendre en compte un ensemble plus large de facteurs lors de l'analyse de la fragmentation des systèmes politiques dans les pays africains.

\begin{table}[H]
\centering
\caption{Nombre effectif de partis pour les pays africains utilisant les indices de Noua. \\ }
\label{tab:NouaIndexes}
\scalebox{0.8}{ 
\begin{tabular}{lccc}
\toprule
\textbf{Pays} & \textbf{Indice Hard} & \textbf{Indice Soft} & \textbf{Moyenne des deux modèles} \\
\midrule
Afrique du Sud           & 11.25      & 10.89      & 11.07               \\ \hline
Algérie                  & 11.61      & 17.79      & 14.7                \\ \hline
Angola                   & 10.73      & 9.87       & 10.3                \\ \hline
Bénin                    & 7.39       & 2          & 4.7                 \\ \hline
Botswana                 & 7.27       & 3.95       & 5.61                \\ \hline
Burkina Faso             & 8.75       & 2.84       & 5.79                \\ \hline
Burundi                  & 5.94       & 2          & 3.97                \\ \hline
Côte d'Ivoire            & 9.22       & 3.52       & 6.37                \\ \hline
Cameroun                 & 9.51       & 4.41       & 6.96                \\ \hline
Cap-Vert                 & 2          & 2          & 2                   \\ \hline
Centrafrique             & 8.07       & 4.34       & 6.21                \\ \hline
Comores                  & 2          & 2          & 2                   \\ \hline
Congo                    & 7.66       & 2.53       & 5.09                \\ \hline
Djibouti                 & 3.29       & 2          & 2.65                \\ \hline
Egypte                   & 11.65      & 11.73      & 11.69               \\ \hline
Eswatini                 & 3.06       & 2          & 2.53                \\ \hline
Ethiopie                 & 11.86      & 13.04      & 12.45               \\ \hline
Érythrée                 & 6.01       & 2          & 4                   \\ \hline
Gabon                    & 6.5        & 2          & 4.25                \\ \hline
Gambie                   & 3.42       & 2          & 2.71                \\ \hline
Ghana                    & 8.99       & 3.08       & 6.04                \\ \hline
Guinée                   & 8.17       & 2.27       & 5.22                \\ \hline
Guinee-Bissau            & 4.35       & 2          & 3.17                \\ \hline
Guinée équatoriale       & 3.95       & 2          & 2.98                \\ \hline
Kenya                    & 10.38      & 6.3        & 8.34                \\ \hline
Lesotho                  & 4.25       & 2          & 3.12                \\ \hline
Liberia                  & 6.42       & 2          & 4.21                \\ \hline
Libye                    & 9.46       & 11.95      & 10.71               \\ \hline
Madagascar               & 9.82       & 5.27       & 7.54                \\ \hline
Malawi                   & 7.82       & 2          & 4.91                \\ \hline
Mali                     & 10.29      & 9.25       & 9.77                \\ \hline
Mauritanie               & 8.55       & 7.03       & 7.79                \\ \hline
Maurice                  & 2          & 2          & 2                   \\ \hline
Maroc                    & 10.2       & 6.38       & 8.29                \\ \hline
Mozambique               & 10.2       & 6.8        & 8.5                 \\ \hline
Namibie                  & 7.8        & 5.59       & 6.7                 \\ \hline
Niger                    & 10.41      & 9.54       & 9.97                \\ \hline
Nigeria                  & 12.26      & 16.27      & 14.27               \\ \hline
RD Congo                 & 12.42      & 20.21      & 16.32               \\ \hline
Rwanda                   & 5.91       & 2          & 3.96                \\ \hline
Sao Tomé-et-Principe     & 2          & 2          & 2                   \\ \hline
Sénégal                  & 8.18       & 2.11       & 5.14                \\ \hline
Seychelles               & 2          & 2          & 2                   \\ \hline
Sierra Leone             & 6.41       & 2          & 4.21                \\ \hline
Somalie                  & 9.37       & 5.03       & 7.2                 \\ \hline
Soudan                   & 11.44      & 14.53      & 12.98               \\ \hline
Soudan du Sud            & 8.91       & 4.77       & 6.84                \\ \hline
Tanzanie                 & 11.05      & 9.22       & 10.14               \\ \hline
Tchad                    & 10.12      & 9.34       & 9.73                \\ \hline
Togo                     & 6.27       & 2          & 4.13                \\ \hline
Tunisie                  & 7.6        & 2          & 4.8                 \\ \hline
Uganda                   & 9.37       & 3.77       & 6.57                \\ \hline
Zambie                   & 9.65       & 5.9        & 7.78                \\ \hline
Zimbabwe                 & 8.76       & 3.31       & 6.04                \\ \hline
\bottomrule
\end{tabular}
}
\end{table}

\subsubsection{Comparaison des modèles Hard et Soft}

La comparaison des deux modèles montre clairement que, bien qu'ils présentent tous deux une tendance généralement croissante, l'indice Soft affiche un NEP supérieur à celui du modèle \textit{Hard}. D'après les résultats obtenus avec les deux approches, l'indice Hard se révèle être un indicateur plus nuancé et réactif aux facteurs liés à la population et à la superficie, capturant une plage de valeurs plus étendue entre les pays. Cela suggère qu'il intègre efficacement des aspects subtils de ces variables. En revanche, l'indice Soft agit comme un indicateur basé sur un seuil, mettant en exergue les pays à forte population et à vaste superficie. Autrement dit, il suggère qu'une plus grande superficie et une population plus importante devraient être associées à un nombre de partis politiques plus élevé par rapport aux pays de taille réduite ou modérée. En termes d'observations spécifiques, la République démocratique du Congo présente le nombre effectif de partis le plus élevé, suivie par le Nigeria, l'Algérie, le Soudan, l'Éthiopie, l'Égypte et l'Afrique du Sud, qui affichent également des valeurs du NEP relativement élevées. À l'inverse, les Seychelles, Sao Tomé-et-Principe, les Comores, Maurice et le Cap-Vert se distinguent par leurs faibles valeurs du NEP.\\

En résumé, d'après les deux modèles, nous pouvons catégoriser le paysage politique en Afrique en deux grands groupes : le système bipartite et le système multipartite. Nous suggérons que les Seychelles, Sao Tomé-et-Principe, les Comores, Maurice et le Cap-Vert devraient adopter un système bipartite. De plus, sur la base des valeurs moyennes issues des deux modèles, nous proposons qu'un pays ne devrait pas présenter un NEP supérieur à 20. Par ailleurs, le paysage politique peut être divisé en trois niveaux de nombre effectif de partis, selon les résultats des deux modèles :

\begin{itemize}
    \item \textbf{Faible} : valeurs du NEP inférieures ou égales à 5 (\( NEP \leq 5 \)).
    \item \textbf{Moyen} : valeurs du NEP comprises entre plus de 5 et jusqu'à 10 (\( 5 < NEP \leq 10 \)).
    \item \textbf{Élevé} : valeurs du NEP supérieures à 10 (\( NEP > 10 \)).
\end{itemize}

\subsubsection{Dynamique temporelle : Impact de l'évolution de la population et de la superficie sur le NEP}

Les Figures \ref{fig:hard-year-compare} et \ref{fig:soft-year-compare} illustrent le nombre effectif de partis dans divers pays africains pour les années 2014 et 2023. Ces comparaisons visent à révéler comment les tendances des modèles hard et soft évoluent sur une période de dix ans, en considérant une superficie invariable et la croissance de la population. Dans la Figure \ref{fig:hard-year-compare}, l'indice Hard montre que la plupart des pays ont connu une légère augmentation du nombre effectif de partis entre 2014 et 2023, soulignant la stabilité de ce modèle. Cependant, certains pays, tels que les Seychelles, Sao Tomé-et-Principe, les Comores, Maurice et le Cap-Vert, ont maintenu de manière constante un NEP de 2, sans variation sur la décennie, ce qui suggère que la croissance démographique a un impact minimal sur le paysage politique dans ces nations. En revanche, la Figure \ref{fig:soft-year-compare} démontre une distinction claire entre deux groupes de pays. Un premier groupe, incluant les Seychelles, Sao Tomé-et-Principe, les Comores, Maurice, le Cap-Vert, la Gambie, Eswatini, Djibouti, le Rwanda, la Guinée équatoriale, le Lesotho, la Guinée-Bissau, le Togo, la Sierra Leone, le Libéria, le Bénin, l'Érythrée, le Malawi et la Tunisie, a maintenu un NEP constant de 2 sur la décennie. Un second groupe présente des valeurs du NEP supérieures à 2 et une plus grande variabilité, avec une légère augmentation globale de 2014 à 2023. Ce schéma suggère que le modèle soft démontre également une stabilité sur une période de dix ans.

\begin{landscape}
\begin{figure}[H]
    \includegraphics[scale=0.7]{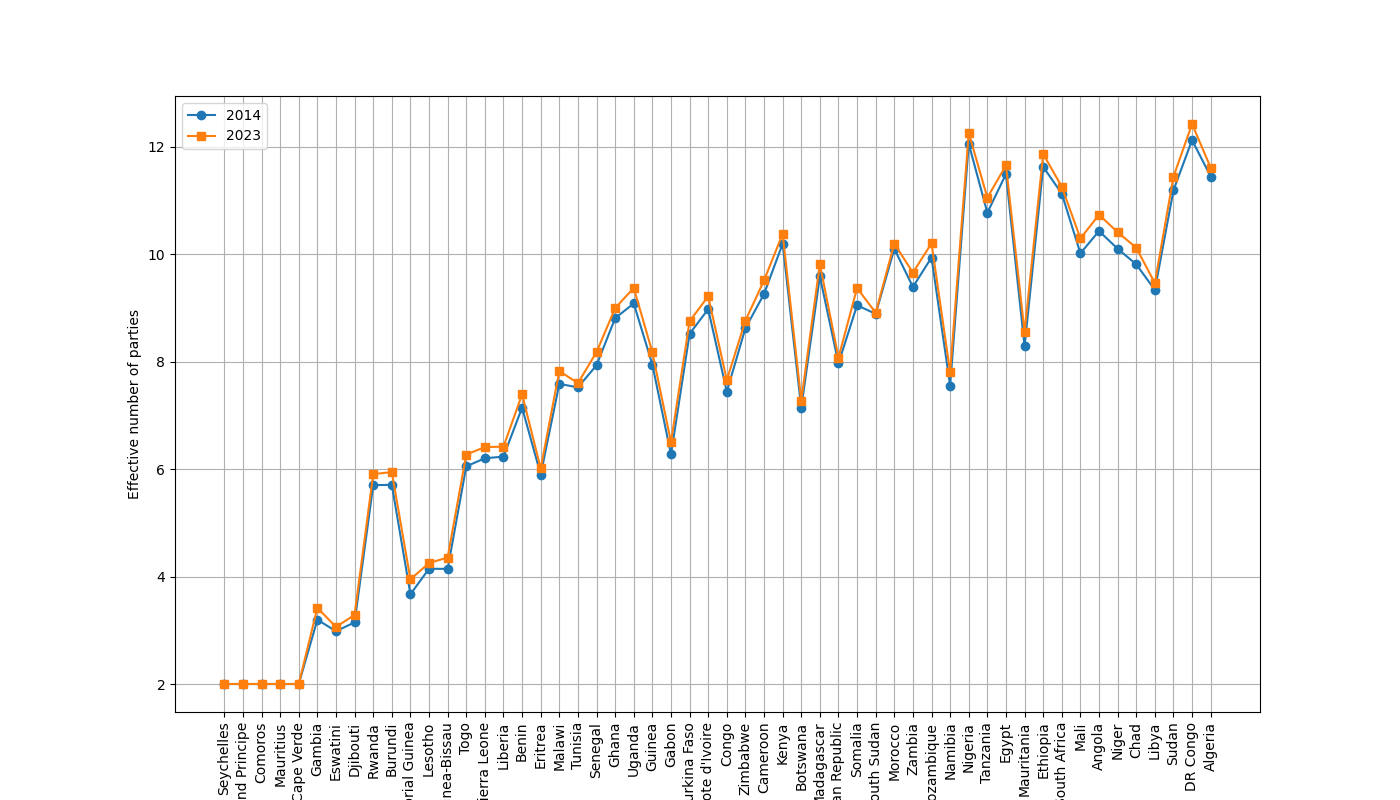}
    \caption{Tendances des systèmes de partis politiques en Afrique utilisant l'indice Hard : Analyse de l'évolution du nombre effectif de partis politiques de 2014 à 2023.}\label{fig:hard-year-compare}
\end{figure}

\begin{figure}[H]
    \includegraphics[scale=0.7]{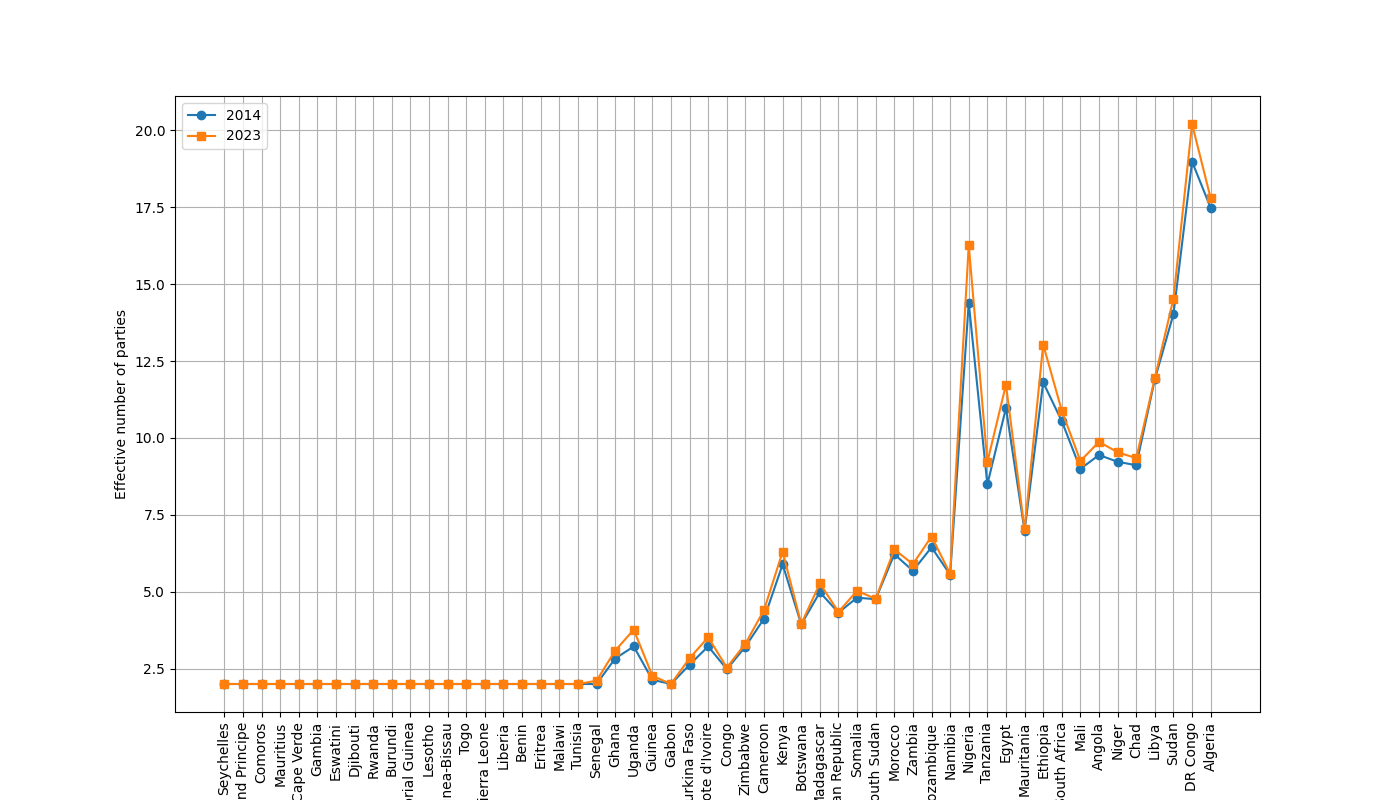}
    \caption{Tendances des systèmes de partis politiques en Afrique utilisant l'indice Soft : Analyse de l'évolution du nombre effectif de partis politiques de 2014 à 2023.} \label{fig:soft-year-compare}
\end{figure}
\end{landscape}

\section{Flexibilité de notre modèle Soft}
Notre modèle Soft fait preuve d'une grande flexibilité, permettant à chaque pays de définir son contexte en ajustant les valeurs de \( k_1 \) et \( k_2 \) afin d'obtenir le nombre effectif de partis correspondant à sa situation. Les résultats présentés ici n'incluent pas les valeurs optimales du NEP pour chaque pays, telles que présentées précédemment, mais mettent plutôt en lumière les dynamiques politiques spécifiques à chaque pays. Dans cette section, nous nous concentrons sur certains pays d'Afrique de l'Ouest et d'Afrique centrale, régions caractérisées par une fragmentation politique particulièrement élevée. Les valeurs de \( k_1 \) et \( k_2 \) utilisées sont calculées au niveau continental, bien qu'elles puissent également être ajustées pour des échelles régionales ou communautaires selon les besoins de l'analyse. Cette approche flexible permet une analyse nuancée du paysage politique, en tenant compte des spécificités locales tout en fournissant une base de comparaison au niveau continental. La Table \ref{tab:k1k2_values} ci-dessous présente différentes valeurs de \( k_1 \) et \( k_2 \), calculées à partir de statistiques descriptives (moyenne et quartiles) des données de population et de superficie des pays africains. Les quartiles sont des mesures statistiques qui divisent un ensemble de données en quatre parts égales. Ils permettent de comprendre la répartition des données en identifiant des points clés. Le \textbf{Q1}, également appelé premier quartile, représente la valeur en dessous de laquelle se situe 25\% des données. Le \textbf{Q2}, ou médiane, divise les données en deux parts égales, indiquant la valeur centrale prédite par le modèle. Le \textbf{Q3}, ou troisième quartile, représente la valeur en dessous de laquelle se situe 75\% des données. Ces valeurs aident à comprendre comment les valeurs de \( k_1 \) et \( k_2 \) influencent la distribution du NEP. Q1div2, Q2div2, Q3div2 représentent respectivement leurs divisions respectives par 2.

\begin{table}[H]
\centering
\caption{Différentes valeurs de \( k_1 \) et \( k_2 \)}
\label{tab:k1k2_values}
\begin{tabular}{|c|c|c|}
\hline
Statistiques   & \( k_1 \)          & \( k_2 \)        \\ \hline
Moyenne     & 26123118.10 & 511180.71 \\ \hline
Q1(25\%) & 2525592.25  & 28680.00  \\ \hline
Q2(50\%) & 14152176.50 & 269800.00 \\ \hline
Q3(75\%) & 31957274.50 & 814062.50 \\ \hline
\end{tabular}
\end{table}

Les résultats sont présentés dans les sections suivantes, où, pour chaque pays, chaque cellule de la carte thermique représente le nombre effectif de partis pour une combinaison spécifique de valeurs de \( k_1 \) et \( k_2 \). 

\subsection{Nombre effectif de partis politiques en Afrique de l'Ouest}

\subsubsection{Cas des pays de l'AES}

\begin{figure}[H]
\centering
\includegraphics[scale=0.85]{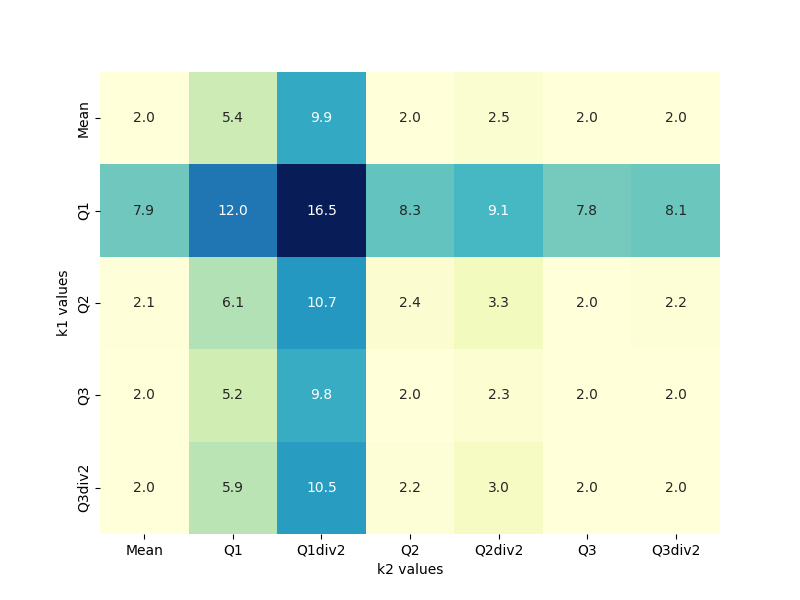}
\caption{Valeurs du Nombre Effectif de Partis pour le Burkina Faso, illustrant l'impact de différentes combinaisons de $k_1$ et $k_2$ à travers diverses mesures statistiques. La carte thermique affiche une matrice des valeurs du NEP. Chaque cellule de la carte thermique représente le NEP pour un appariement spécifique de $k_1$ et $k_2$, où ces paramètres prennent des valeurs dérivées de la moyenne (Mean), des quartiles (Q1, Q2, Q3), et leurs divisions respectives par 2 (Q1div2, Q2div2, Q3div2). L'intensité de la couleur représente l'amplitude des valeurs du NEP, avec des nuances plus foncées indiquant des valeurs plus élevées.}
\label{fig:hfaso}
\end{figure}

La Figure \ref{fig:hfaso} illustre les résultats pour le Burkina Faso, révélant des valeurs du NEP qui varient considérablement, allant d'un minimum de 2.0 à un maximum de 16.5. Les valeurs de NEP les plus faibles, proches du seuil de 2, se manifestent principalement lorsque le paramètre $k_1$, associé à la population, prend des valeurs relativement élevées (Moyenne, Q3, Q3div2). Dans ces configurations du modèle, l'influence de la population sur la fragmentation du système de partis est moindre. Il est intéressant de noter que dans ces scénarios de $k_1$ élevé, les variations des valeurs de $k_2$ (superficie) ont un impact limité sur le NEP, qui reste stable autour de 2.0. Cela suggère que si la population est considérée comme un facteur de fragmentation moins important (selon les valeurs de $k_1$ choisies), le système de partis au Burkina Faso tend vers un modèle bipartite, avec peu de sensibilité à la contribution de la superficie. Des valeurs du NEP modérées, se situant approximativement entre 3 et 6, apparaissent dans des configurations où $k_1$ prend des valeurs intermédiaires (Q2, Q2div2) et où $k_2$ varie. Dans ces cas, l'influence conjointe de la population et de la superficie conduit à une fragmentation plus marquée qu'un système bipartite, mais sans atteindre les niveaux extrêmes observés dans d'autres scénarios. Ces résultats suggèrent que selon certaines pondérations de la population et de la superficie, le Burkina Faso pourrait évoluer vers un système multipartite limité, avec un nombre effectif de partis légèrement supérieur à deux. Les valeurs du NEP les plus élevées, atteignant jusqu'à 16.5, se produisent lorsque le modèle attribue une forte influence à la fois à la population (avec une valeur de $k_1$ au premier quartile, Q1) et à la superficie (avec une valeur de $k_2$ relativement faible, Q1 ou Q1div2). Ce scénario extrême indique que si les caractéristiques démographiques et géographiques du Burkina Faso sont interprétées comme des facteurs majeurs de fragmentation (à travers le choix de ces valeurs spécifiques de $k_1$ et $k_2$), le pays pourrait potentiellement avoir un système multipartite très fragmenté. Cette configuration met en lumière le potentiel de dispersion des forces politiques en de nombreux partis effectifs.

\begin{figure}[H]
\centering
\includegraphics[scale=0.85]{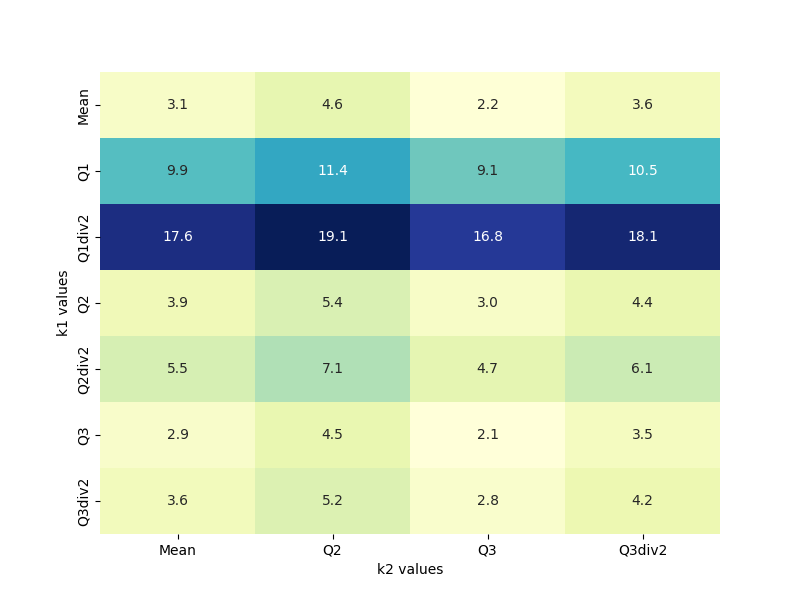}
\caption{Valeurs du Nombre Effectif de Partis pour le Mali, illustrant l'impact de différentes combinaisons de $k_1$ et $k_2$ à travers diverses mesures statistiques. La carte thermique affiche une matrice des valeurs du NEP. Chaque cellule de la carte thermique représente le NEP pour un appariement spécifique de $k_1$ et $k_2$, où ces paramètres prennent des valeurs dérivées de la moyenne, des quartiles (Q1, Q2, Q3), et leurs divisions respectives par 2 (Q1div2, Q2div2, Q3div2). L'intensité de la couleur représente l'amplitude des valeurs du NEP, avec des nuances plus foncées indiquant des valeurs plus élevées.}
\label{fig:hmali}
\end{figure}

La Figure \ref{fig:hmali} révèle le NEP du Mali, avec des valeurs faibles (environ 2.1 à 7.1) généralement observées lorsque $k_1$ est fixé à la Moyenne, Q2, Q2div2, Q3 ou Q3div2, quelle que soit la valeur de $k_2$, suggérant un système politique moins fragmenté dans ces conditions. Il est intéressant de constater que dans ces contextes de $k_1$ élevé, les variations des valeurs de $k_2$ n'entraînent qu'une faible incidence sur le NEP. Les valeurs les plus élevées du NEP (16.8 à 19.1) se concentrent lorsque $k_1$ est fixé à Q1div2, indiquant une forte influence de la population sur la fragmentation. Le pic du NEP (19.1) est atteint avec $k_1$ à Q1div2 et $k_2$ à Q2, soulignant une interaction notable entre la moitié du premier quartile de la population et la médiane de la superficie dans la configuration du paysage malien. De plus, un niveau de fragmentation modéré (NEP entre 9.1 et 11.4) est noté lorsque $k_1$ est fixé à Q1, se situant entre les extrêmes observés.\\

\begin{figure}[H]
\centering
\includegraphics[scale=0.85]{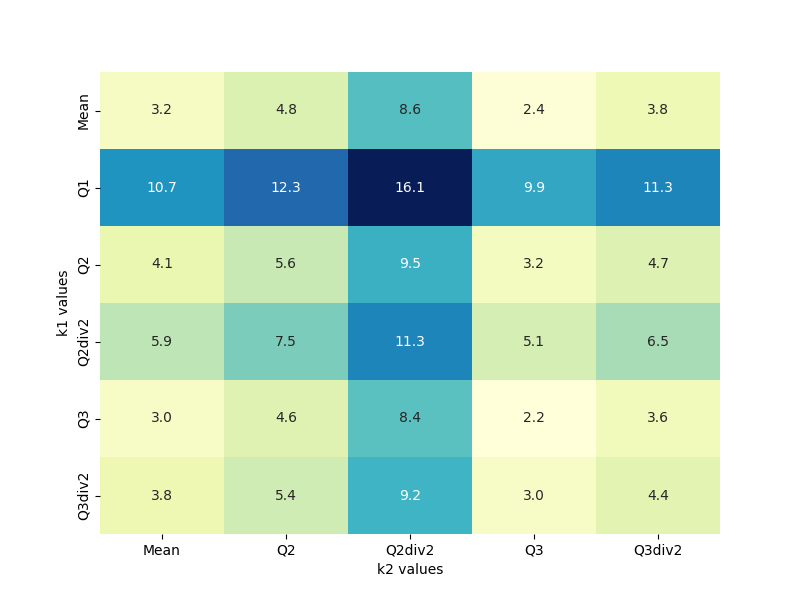}
\caption{Valeurs du Nombre Effectif de Partis pour le Niger, illustrant l'impact de différentes combinaisons de $k_1$ et $k_2$ à travers diverses mesures statistiques. La carte thermique affiche une matrice des valeurs du NEP. Chaque cellule de la carte thermique représente le NEP pour un appariement spécifique de $k_1$ et $k_2$, où ces paramètres prennent des valeurs dérivées de la moyenne, des quartiles (Q1, Q2, Q3), et leurs divisions respectives par 2 (Q2div2, Q3div2). L'intensité de la couleur représente l'amplitude des valeurs du NEP, avec des nuances plus foncées indiquant des valeurs plus élevées.}
\label{fig:hniger}
\end{figure}

La Figure \ref{fig:hniger} pour le Niger illustre la relation entre différentes combinaisons de \( k_1 \) et \( k_2 \) et leur impact sur le nombre effectif de partis. En général,  les configurations caractérisées par des valeurs élevées du paramètre $k_1$ (correspondant à la moyenne , au troisième quartile ou au troisième quartile divisé par deux) produisent invariablement un NEP minimal (inférieur à 6), indépendamment des variations du paramètre $k_2$. Cette tendance suggère une propension à une fragmentation minimale dans les contextes où la population atteint une certaine masse critique. La valeur minimale de 2.2 est atteinte lorsque les deux paramètres sont fixés à Q3. À l'opposé du spectre, les données révèlent un phénomène de fragmentation maximale (NEP = 16.1) lorsque \( k_1 \) est calibré au premier quartile de distribution et \( k_2 \) à la médiane divisée par deux, ce qui démontre un niveau élevé de fragmentation du paysage niégrien lorsque Q1 est combiné avec Q2div2. Il convient également de noter les valeurs modérées du NEP, généralement comprises entre 8 et 12, sont observées lorsque \( k_1 \) est fixé à Q1 en association avec d'autres valeurs de \( k_2 \) ou à \( k_2 \) fixe avec des valeurs différent de $k_1$. Ces valeurs modérées suggèrent que le premier quartile contribue à un système  plus fragmenté, bien que moins extrême que la combinaison Q1 et Q2div2.

\hfill \break

\subsubsection{Cas des pays de la CEDEAO}
Ici, nous présentons les résultats pour trois pays à savoir le Bénin, le Sénégal et le Togo.\\

\begin{figure}[H]
\centering
\includegraphics[scale=0.85]{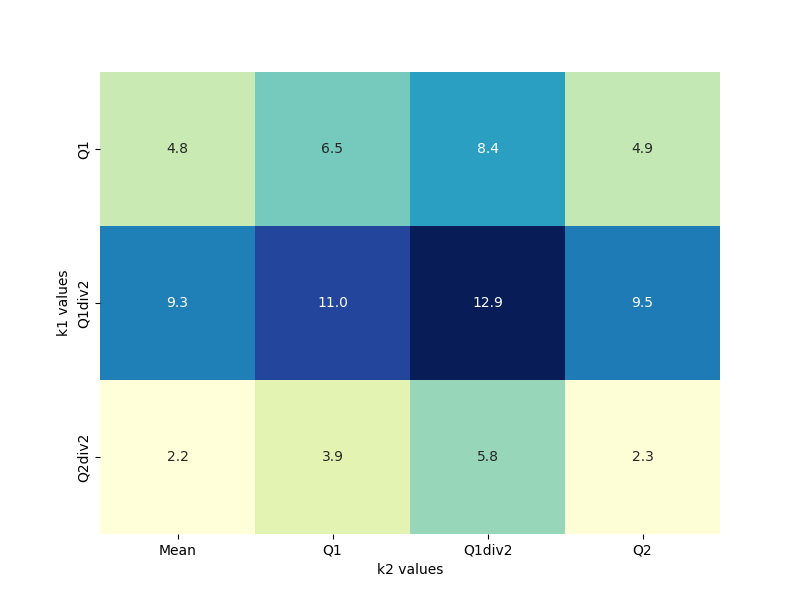}
\caption{Valeurs du Nombre Effectif de Partis pour le Bénin, illustrant l'impact de différentes combinaisons de $k_1$ et $k_2$ à travers diverses mesures statistiques. La carte thermique affiche une matrice des valeurs du NEP. Chaque cellule de la carte thermique représente le NEP pour un appariement spécifique de $k_1$ et $k_2$, où ces paramètres prennent des valeurs dérivées de la moyenne, des quartiles (Q1, Q2, Q3), et leurs divisions respectives par 2 (Q1div2, Q2div2, Q3div2). L'intensité de la couleur représente l'amplitude des valeurs du NEP, avec des nuances plus foncées indiquant des valeurs plus élevées.}
\label{fig:benin}
\end{figure}

Concernant le Bénin, la Figure \ref{fig:benin} illustre l'influence combinée de la population ($k_1$) et de la superficie ($k_2$) sur le NEP. Un NEP faible est principalement observé lorsque la population est relativement faible (Q2div2), avec des valeurs allant de 2.2 à 5.8, suggérant une fragmentation limitée dans ces conditions. Une fragmentation modérée (NEP entre 4.8 et 8.4) émerge lorsque la population se situe au premier quartile (Q1), indiquant une influence notable de ce paramètre sur la division politique. Le NEP le plus élevé (jusqu'à 12.9) est atteint avec une population très faible (Q1div2) combinée à une superficie faile (Q1div2), signalant une forte fragmentation pour le paysage béninois. On note que même avec une population faible (Q1), l'interaction avec différentes valeurs de superficie module le niveau de fragmentation, sans atteindre les extrêmes observés avec une population encore plus faible (Q1div2).\\

\begin{figure}[H]
\centering
\includegraphics[scale=0.85]{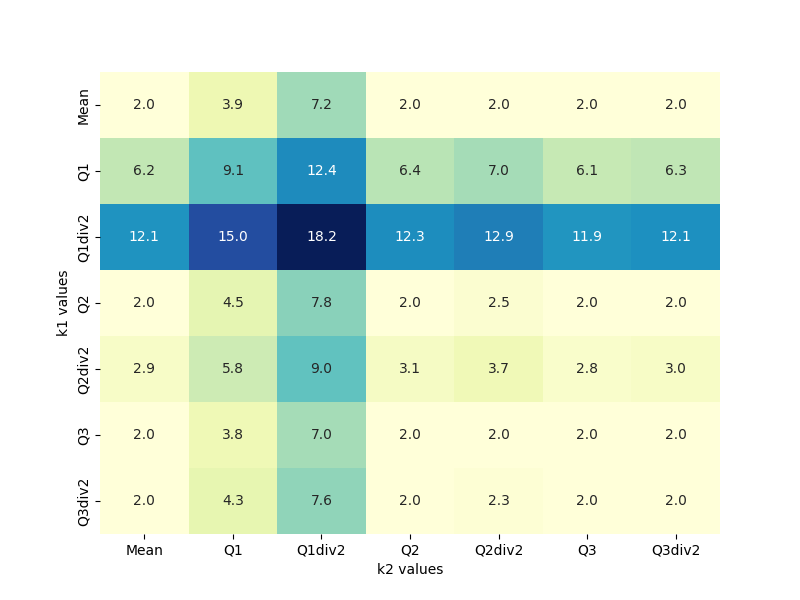}
\caption{Valeurs du Nombre Effectif de Partis pour le Sénégal, illustrant l'impact de différentes combinaisons de $k_1$ et $k_2$ à travers diverses mesures statistiques. La carte thermique affiche une matrice des valeurs du NEP. Chaque cellule de la carte thermique représente le NEP pour un appariement spécifique de $k_1$ et $k_2$, où ces paramètres prennent des valeurs dérivées de la moyenne, des quartiles (Q1, Q2, Q3), et leurs divisions respectives par 2 (Q1div2, Q2div2, Q3div2). L'intensité de la couleur représente l'amplitude des valeurs du NEP, avec des nuances plus foncées indiquant des valeurs plus élevées.}
\label{fig:senegal}
\end{figure}

Le paysage du Sénégal est illustré par la Figure \ref{fig:senegal} ou les valeurs du NEP varient de manière étendue, allant d'un minimum de 2.0 jusqu'à un pic de 18.2. Cette variation illustre que, selon la pondération accordée à la population et à la superficie, le modèle peut prédire pour le Sénégal soit un système politique très concentré autour de deux pôles, soit un système potentiellement très fragmenté. Le plancher de 2.0, fréquemment observé, indique des combinaisons de $k_1$ et $k_2$ où la contribution combinée de la population et de la superficie est jugée trop faible par le modèle pour soutenir un plus grand nombre de partis. Les dynamiques spécifiques de la carte thermique montrent que les valeurs de NEP les plus élevées pour le Sénégal sont atteintes lorsque les coefficients $k_1$ et $k_2$ sont les plus faibles (par exemple, les combinaisons impliquant "Q1div2"). Cela signifie que si l'on considère que la population et la superficie exercent une influence très forte (dénominateurs $k_1, k_2$ petits), le Sénégal est modélisé comme ayant un nombre élevé de partis effectifs. Inversement, lorsque $k_1$ et $k_2$ prennent des valeurs plus grandes (comme les valeurs "Mean" ou "Q3"), diminuant l'impact proportionnel de la population et de la superficie, le NEP tend vers le minimum de 2.0. Cela suggère que si les facteurs démographiques et géographiques nécessitent un seuil plus élevé pour impacter le NEP, le modèle converge vers une structure bipartisane pour le pays.


\begin{figure}[H]
\centering \includegraphics[scale=0.85]{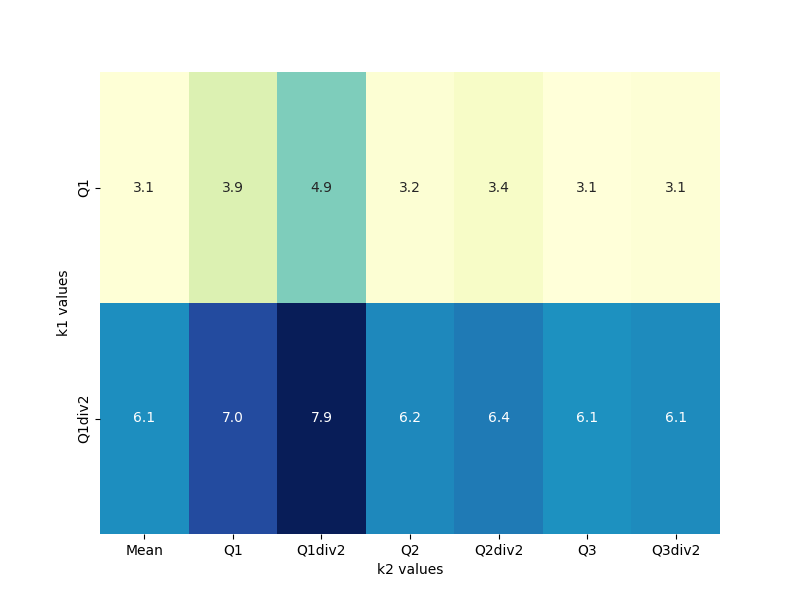}
\caption{Valeurs du Nombre Effectif de Partis pour le Togo, illustrant l'impact de différentes combinaisons de $k_1$ et $k_2$ à travers diverses mesures statistiques. La carte thermique affiche une matrice des valeurs du NEP. Chaque cellule de la carte thermique représente le NEP pour un appariement spécifique de $k_1$ et $k_2$, où ces paramètres prennent des valeurs dérivées de la moyenne, des quartiles (Q1, Q2, Q3), et leurs divisions respectives par 2 (Q1div2, Q2div2, Q3div2). L'intensité de la couleur représente l'amplitude des valeurs du NEP, avec des nuances plus foncées indiquant des valeurs plus élevées.}
\label{fig:togo}
\end{figure}

La figure \ref{fig:togo} illustre l'impact de différentes combinaisons de \( k_1 \) et \( k_2 \) sur le NEP au Togo. Lorsque \( k_1 \) est fixé à Q1, les valeurs du NEP restent constamment faibles, variant de 3.1 à 4.9, quelle que soit la valeur de \( k_2 \). La valeur minimale du NEP au Togo est observée lorsque \( k_2 \) est fixé à la Moyenne, à Q3 ou à Q3div2, ce qui suggère une fragmentation minimale du système politique lorsque la population est faible \( k_1 =Q1 \). En revanche, lorsque \( k_1 \) est fixé à Q1div2, les valeurs du NEP sont généralement plus élevées, variant de 6.1 à 7.9. La valeur maximale du NEP (7.9) survient lorsque \( k_2 \) est fixé à Q1div2, indiquant un degré relativement élevé de fragmentation du système togolais lorsque \( k_1 \) et \( k_2 \) sont tous deux fixés à Q1div2. Des valeurs du NEP modérées, comprises entre 6.1 et 6.4, apparaissent avec diverses autres valeurs de \( k_2 \) (Moyenne, Q2, Q2div2, Q3, Q3div2), indiquant que Q1div2 contribue de manière plus significative à la fragmentation politique togolaise que Q1.

\hfill \break

\subsection{Nombre effectif de partis politiques en Afrique centrale}

Dans cette section, nous analysons le nombre effectif de partis dans trois pays membres de la Communauté économique des États de l'Afrique centrale : le Cameroun, le Tchad et la République démocratique du Congo.\\

\begin{figure}[H]
\centering
\includegraphics[scale=0.85]{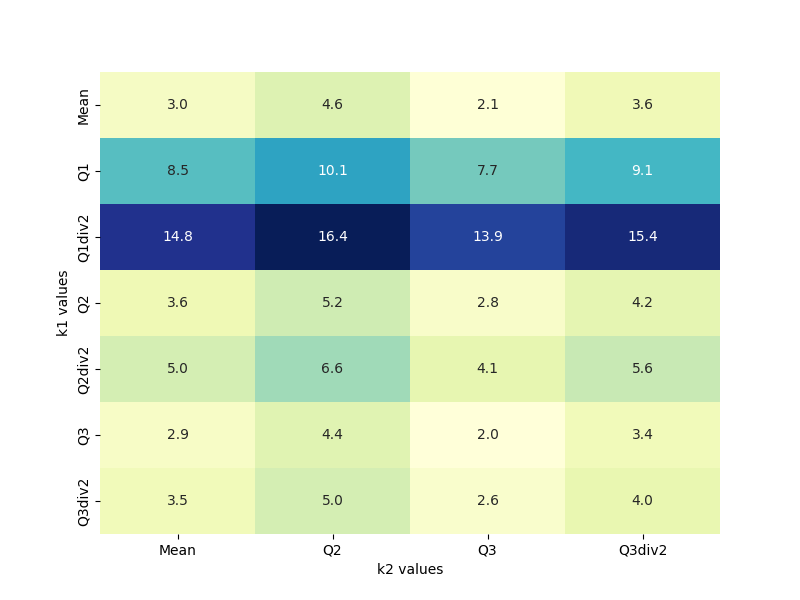}
\caption{Valeurs du nombre effectif de partis pour le Tchad, illustrant l'impact de différentes combinaisons de \( k_1 \) et \( k_2 \) à travers diverses mesures statistiques. La carte thermique affiche une matrice de valeurs du NEP. Chaque cellule de la carte représente le NEP pour une paire spécifique de \( k_1 \) et \( k_2 \), où ces paramètres prennent des valeurs dérivées de la moyenne, des quartiles (Q1, Q2, Q3) et de leurs divisions par 2 (Q1div2, Q2div2, Q3div2). L'intensité des couleurs représente la magnitude des valeurs du NEP, des teintes plus foncées indiquant des valeurs plus élevées.}
\label{fig:chad}
\end{figure}

La Figure \ref{fig:chad} présente le cas du Tchad. On y constate que, lorsque \( k_1 \) est fixé à la Moyenne, à Q2, à Q3 ou à Q3div2, les valeurs du NEP sont nettement faibles, généralement inférieures à 6.6. Cela suggère un système relativement consolidé dans ces conditions. En particulier, les valeurs les plus faibles du NEP sont observées lorsque \( k_1 \) ou \( k_2 \) valent Q3, l'absolu minimum étant atteint lorsque \( k_1 \) et \( k_2 \) sont fixés à Q3, avec un NEP de 2.0.  À l'inverse, une augmentation nette des valeurs du NEP est constatée lorsque \( k_1 \) est fixé à Q1 ou à Q1div2. Les valeurs du NEP les plus élevées, comprises entre 13.9 et 16.4, se produisent lorsque \( k_1 \) est fixé à Q1div2 indépendamment de $k_2$. La valeur maximale du NEP (16.4) apparaît précisément lorsque \( k_1 \) est fixé à Q1div2 et \( k_2 \) à Q2, ce qui indique que la combinaison de la moitié du premier quartile avec le deuxième quartile est fortement associée à une fragmentation accrue du système politique au Tchad. Lorsque \( k_1 \) est fixé à Q1, en fonction de la valeur de \( k_2 \), le NEP varie de 7,7 à 10,1 démontrant une fragmentation modérée.

\begin{figure}[H]
\centering
\includegraphics[scale=0.85]{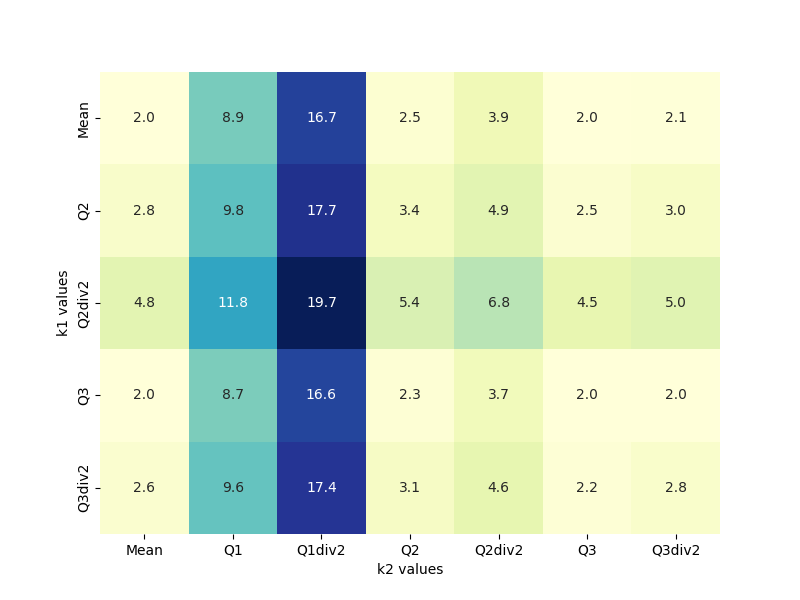}
\caption{Valeurs du nombre effectif de partis pour le Cameroun, illustrant l'impact de différentes combinaisons de \( k_1 \) et \( k_2 \) à travers diverses mesures statistiques. La carte thermique affiche une matrice de valeurs du NEP. Chaque cellule de la carte représente le NEP pour une paire spécifique de \( k_1 \) et \( k_2 \), où ces paramètres prennent des valeurs dérivées de la moyenne, des quartiles (Q1, Q2, Q3) et de leurs divisions par 2 (Q1div2, Q2div2, Q3div2). L'intensité des couleurs représente la magnitude des valeurs du NEP, des teintes plus foncées indiquant des valeurs plus élevées.}
\label{fig:camer}
\end{figure}

Les résultats pour le Cameroun, présentés dans la Figure \ref{fig:camer}, montrent que des valeurs faibles du NEP (généralement inférieures à 7) sont observées lorsque \( k_2 \) est fixé à Q2, Q2div2, Q3 ou Q3div2 et à la Moyenne. Cela implique une certaine stabilité dans la diversité politique. Le minimum de 2.0 est atteint  avec les combinaisons suivant  \( k_1 = mean , k_2=mean \), \( k_1 = Q3 , k_2=mean \),\( k_1 = Q3 , k_2=Q3 \) et \( k_1 = Q3 , k_2=Q3div2 \). Les valeur modérées sont observées lorsqu'on utilise le premier quartile pour la superficie, allant de 8.7 à 11.8 suggérant une diversité politique modérée, avec plusieurs partis ayant une représentation significative. Les valeurs du NEP les plus élevées (allant de 16.6 à 19.7) se regroupent lorsque \( k_2 \) est fixé à Q1div2 avec le maximum atteint avec \( k_1 \) à Q2div2, indiquant un niveau de fragmentation le plus élevé, démontrant une présence significative de partis divers. En somme, la faible superficie (Q1div2 et Q1) est un paramètre qui influence de manière significative la fragmentation très diversifié du système camerounais tandis que une superficie élevée montre une fragmentation relativement moins diversifiée.

\begin{figure}[H]
\centering \includegraphics[scale=0.85]{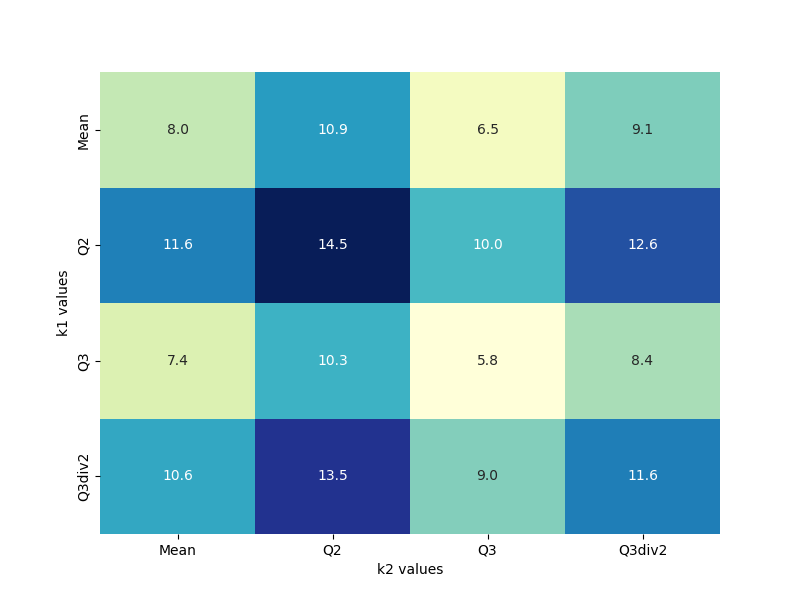}
\caption{Valeurs du nombre effectif de partis pour le RD Congo, illustrant l'impact de différentes combinaisons de \( k_1 \) et \( k_2 \) à travers diverses mesures statistiques. La carte thermique affiche une matrice de valeurs du NEP. Chaque cellule de la carte représente le NEP pour une paire spécifique de \( k_1 \) et \( k_2 \), où ces paramètres prennent des valeurs dérivées de la moyenne, des quartiles (Q1, Q2, Q3) et de leurs divisions par 2 (Q1div2, Q2div2, Q3div2). L'intensité des couleurs représente la magnitude des valeurs du NEP, des teintes plus foncées indiquant des valeurs plus élevées.}
\label{fig:drcongo}
\end{figure}

Les résultats pour la République démocratique du Congo, présentés dans la Figure \ref{fig:drcongo}, révèlent une distribution hétérogène du Nombre Effectif de Partis, avec des valeurs variant significativement selon les paramètres utilisés. La valeur la plus faible ($5.8$) est observée lorsque \( k_1 \) et \( k_2 \)  sont tous deux fixés à Q3, tandis que la valeur maximale ($14.5$) est atteinte avec \( k_1 \) et \( k_2 \) fixés à Q2. La distribution des valeurs dans la carte thermique montre une tendance à la diminution du NEP lorsque \( k_1 \) ou \( k_2 \) est fixé à Q3, suggérant une concentration relative du pouvoir politique dans ces configurations. À l'inverse, les combinaisons impliquant Q2 produisent généralement des valeurs plus élevées, indiquant une fragmentation accrue. Lorsque \( k_1 \) est fixé à Q3div2, le NEP fluctue entre $9.0$ et $13.5$, démontrant que l'interaction entre la médiane et la moitié du troisième quartile joue un rôle déterminant dans la structuration de la diversité et de la compétitivité des partis politiques. L'analyse comparative confirme que toutes les valeurs du NEP restent supérieures à $5$, seuil dépassant la caractéristique des systèmes bipartites, attestant de la nature multipartite du système politique congolais. Cette caractéristique distingue la RDC de systèmes plus fortement bipolaires observés ailleurs. La configuration optimale pour capturer la concentration du pouvoir politique semble être celle où \( k_1 \) et \( k_2 \) sont fixés à Q3, mais même dans ce cas, le système demeure fondamentalement multipartite.

\section{Évaluation comparative de notre méthode face aux approches existantes}
Nous comparons nos modèles avec des méthodes existantes, incluant l'indice Laakso \cite{Laakso1979}, l'indice Golosov \cite{Golosov2010} et le Seat Product \cite{taagepera2007}. La Table \ref{tab:enp-comparison} présente les résultats comparatifs de nos modèles et de ces méthodes existantes pour prédire le nombre effectif de partis. Les données utilisées pour les trois méthodes existantes proviennent des résultats des plus récentes élections organisées dans le pays concerné. Comme le montre la table, les méthodes existantes ne fournissent pas de valeurs pour des pays tels que l'Érythrée, la Libye, la Somalie et Eswatini en raison de l'absence d'élections ou de données manquantes. \\

Les indices Laakso et Golosov classifient souvent des pays de taille moyenne (par exemple, le Ghana, le Congo) et de grande taille (par exemple, l'Éthiopie, le Soudan, le Soudan du Sud) comme des systèmes à parti unique, une tendance particulièrement marquée pour l'indice Golosov. Cette classification est en contradiction avec les structures politiques officielles de ces pays, aucun d'eux ne se déclarant formellement comme un État à parti unique. Cette discordance soulève d'importantes questions quant à la précision et à la fiabilité de ces indices pour saisir la véritable nature des systèmes partisans, notamment dans des contextes où les dynamiques politiques sont complexes ou la concurrence électorale limitée. De plus, les indices Laakso et Golosov produisent des résultats incohérents, comme en témoignent les valeurs exceptionnellement élevées pour la République démocratique du Congo (34,08 et 36,67 respectivement), qui sont nettement supérieures à la deuxième valeur la plus élevée de 6,94. Par ailleurs, le Seat Product donne généralement des valeurs du NEP inférieures par rapport aux indices de Noua et ne parvient pas à fournir de données en cas de manque d'informations, par exemple dans les pays dépourvus de systèmes électoraux fonctionnels ou de données institutionnelles fiables. Cela met en lumière les limites du recours exclusif à des paramètres institutionnels, tels que les parts de voix et de sièges, lesquels ne reflètent pas toujours pleinement la réalité politique dans certains contextes.\\

En résumé, les méthodes traditionnelles telles que les indices Laakso et Golosov ainsi que le Seat Product sont limitées aux pays disposant de systèmes électoraux fonctionnels et ne peuvent pas capturer avec précision la fragmentation des systèmes politiques dans des contextes où les élections sont absentes ou fortement contrôlées. Ces méthodes reposent sur des données institutionnelles (par exemple, les parts de voix et de sièges), souvent contestées par les groupes d'opposition, ce qui compromet encore davantage leur fiabilité dans certains environnements politiques. En revanche, les modèles de l'indice Noua, fournissent des valeurs du NEP pour tous les pays et même dans les contextes instables ou non démocratiques, démontrant ainsi son avantage d'utilisation et sa robustesse sur les méthodes existantes.

\begin{table}[H]
\centering
\caption{Comparaison du nombre effectif de partis selon nos modèles et les méthodes existantes (Laakso, Golosov et Seat Product) dans les pays africains. Contrairement aux méthodes traditionnelles limitées par les données électorales, les indices de Noua estiment le NEP même en l'absence d’élections, corrigeant des incohérences comme la mauvaise classification de certains pays et des valeurs aberrantes.\\}
\label{tab:enp-comparison}
\scalebox{0.8}{ 
\begin{tabular}{|l|c|c|c|c|c|}
\hline
\textbf{Pays}    & \textbf{Laakso} & \textbf{Golosov} & \textbf{Seat Product} & \textbf{Noua} ( \textbf{ Indice Hard }) & \textbf{Noua} ( \textbf{Indice Soft } ) \\ \hline
Afrique du Sud           & 2.58   & 1.99    & 7.37         & 11.25                   & 10.89                    \\ \hline
Algérie                  & 1.86   & 1.49    & 7.41         & 11.61                   & 17.79                    \\ \hline
Angola                   & 2.2    & 2.05    & 6.04         & 10.73                   & 9.87                     \\ \hline
Bénin                    & 3.48   & 3.12    & 4.78         & 7.39                    & 2                        \\ \hline
Botswana                 & 2.94   & 2.24    & 3.85         & 7.27                    & 3.95                     \\ \hline
Burkina Faso             & 6.45   & 5.15    & 5.03         & 8.75                    & 2.84                     \\ \hline
Burundi                  & 1.79   & 1.45    & 4.93         & 5.94                    & 2                        \\ \hline
Cameroun                 & 1.8    & 1.37    & 5.65         & 9.51                    & 4.41                     \\ \hline
Cap-Vert                 & 2.41   & 2.13    & 4.16         & 2                       & 2                        \\ \hline
Centrafrique             & 6.05   & 5.8     & 5.19         & 8.07                    & 4.34                     \\ \hline
Comores                  & 1.41   & 1.22    & 2.88         & 2                       & 2                        \\ \hline
Congo                    & 1.79   & 1.45    & 5.33         & 7.66                    & 2.53                     \\ \hline
Côte d'Ivoire            & 3.21   & 2.46    & 6.33         & 9.22                    & 3.52                     \\ \hline
Djibouti                 & 1.13   & 1.07    & 4.02         & 3.29                    & 2                        \\ \hline
Egypte                   & 2.98   & 2.28    & 8.4          & 11.65                   & 11.73                    \\ \hline
Érythrée                 & -      & -       & -            & 6.01                    & 2                        \\ \hline
Eswatini                 & -      & -       & -            & 3.06                    & 2                        \\ \hline
Ethiopie                 & 1.07   & 1.04    & 7.77         & 11.86                   & 13.04                    \\ \hline
Gabon                    & 2.07   & 1.62    & 5.23         & 6.5                     & 2                        \\ \hline
Gambie                   & 4.53   & 4.51    & 3.76         & 3.42                    & 2                        \\ \hline
Ghana                    & 1.85   & 1.53    & 6.51         & 8.99                    & 3.08                     \\ \hline
Guinée                   & 2.06   & 1.6     & 4.83         & 8.17                    & 2.27                     \\ \hline
Guinee-Bissau            & 4.05   & 3.35    & 4.67         & 4.35                    & 2                        \\ \hline
Guinée équatoriale       & 2.98   & 2.28    & 8.42         & 3.95                    & 2                        \\ \hline
Kenya                    & 4.03   & 3.34    & 7.04         & 10.38                   & 6.3                      \\ \hline
Lesotho                  & 4.43   & 3.73    & 4.93         & 4.25                    & 2                        \\ \hline
Liberia                  & 6.94   & 6.96    & 4.18         & 6.42                    & 2                        \\ \hline
Libye                    & -      & -       & -            & 9.46                    & 11.95                    \\ \hline
Madagascar               & 2.95   & 2.72    & 5.46         & 9.82                    & 5.27                     \\ \hline
Malawi                   & 4.13   & 3.72    & 5.77         & 7.82                    & 2                        \\ \hline
Mali                     & 5.7    & 4.61    & 5.28         & 10.29                   & 9.25                     \\ \hline
Mauritanie               & 2.6    & 1.97    & 5.6          & 8.55                    & 7.03                     \\ \hline
Maurice                  & 3.42   & 3.17    & 4.12         & 2                       & 2                        \\ \hline
Maroc                    & 5.68   & 5.49    & 7.34         & 10.2                    & 6.38                     \\ \hline
Mozambique               & 1.79   & 1.45    & 6.3          & 10.2                    & 6.8                      \\ \hline
Namibie                  & 2.17   & 1.69    & 4.58         & 7.8                     & 5.59                     \\ \hline
Niger                    & 6.09   & 4.63    & 5.5          & 10.41                   & 9.54                     \\ \hline
Nigeria                  & 2.77   & 2.32    & 7.11         & 12.26                   & 16.27                    \\ \hline
RD Congo                 & 34.08  & 36.67   & 7.81         & 12.42                   & 20.21                    \\ \hline
Rwanda                   & 1.77   & 1.43    & 3.66         & 5.91                    & 2                        \\ \hline
Sao Tomé-et-Principe     & 2.95   & 2.49    & 3.8          & 2                       & 2                        \\ \hline
Sénégal                  & 2.89   & 2.44    & 5.48         & 8.18                    & 2.11                     \\ \hline
Seychelles               & 2.08   & 1.87    & 3.27         & 2                       & 2                        \\ \hline
Sierra Leone             & 2.08   & 1.81    & 5.13         & 6.41                    & 2                        \\ \hline
Somalie                  & -      & -       & -            & 9.37                    & 5.03                     \\ \hline
Soudan du Sud            & 1.11   & 1.06    & 5.54         & 8.91                    & 4.77                     \\ \hline
Soudan                   & 1.72   & 1.4     & 7.52         & 11.44                   & 14.53                    \\ \hline
Tanzanie                 & 1.26   & 1.13    & 7.32         & 11.05                   & 9.22                     \\ \hline
Tchad                    & 4.32   & 3.26    & 5.73         & 10.12                   & 9.34                     \\ \hline
Togo                     & 2.14   & 1.67    & 4.5          & 6.27                    & 2                        \\ \hline
Tunisie                  & 1.47   & 1.25    & 4.61         & 7.6                     & 2                        \\ \hline
Uganda                   & 2.28   & 1.75    & 8.09         & 9.37                    & 3.77                     \\ \hline
Zambie                   & 2.76   & 2.4     & 5.38         & 9.65                    & 5.9                      \\ \hline
Zimbabwe                 & 2.05   & 1.81    & 6.54         & 8.76                    & 3.31                     \\ \hline
\end{tabular}%
}
\end{table}

\section{Conclusion et Perspectives}
Dans cette étude, nous avons introduit une approche structurée visant à rationaliser le paysage politique en Afrique en intégrant des facteurs géographiques et démographiques, notamment la taille de la population et la superficie. Notre modèles, l'indice Noua Hard et Noua Soft, qui sont des une mesure apolitique, calculent efficacement un nombre effectif de partis ajusté à l'aide d'une formule qui équilibre ces facteurs grâce à des constantes ajustables \( k_1 \) et \( k_2 \). Cette méthode garantit un minimum de deux partis politiques, reflétant le besoin universel d'opposition. La flexibilité du modèle Soft permet d'ajuster les paramètres clés (\( k_1 \) et \( k_2 \)) pour prédire le NEP dans divers pays africains. Les résultats ont démontré que notre modèle capture efficacement la diversité des paysages politiques à travers le continent, avec des valeurs du NEP plus élevées observées dans les pays plus vastes et plus peuplés, et des valeurs plus faibles dans les nations de moindre envergure. Notre analyse a révélé une variabilité significative du NEP entre les différents pays africains, soulignant l'impact des facteurs géographiques et démographiques sur la fragmentation politique. Les cartes thermiques pour divers pays ont montré comment différentes combinaisons de \( k_1 \) et \( k_2 \) influencent le NEP, les valeurs les plus élevées apparaissant souvent lorsque ces paramètres sont alignés sur des quartiles spécifiques.\\

Notre travail établit une base solide pour alimenter les discussions concernant la rationalisation du nombre de partis politiques en Afrique. Bien que ce travail ne préconise pas la manière dont cette réduction devrait s'opérer ni quels partis devraient perdurer, il ouvre la voie à l'exploration des implications politiques d'un nombre effectif de partis optimal pour chaque pays. De telles perspectives pourraient informer des stratégies de gouvernance visant à équilibrer la représentation politique et la stabilité. Dans nos travaux futurs, nous explorerons les mécanismes permettant d'atteindre cette rationalisation en proposant des scénarios de fusion et de coopération entre les partis politiques existants dans chaque pays. Nous sommes convaincus que favoriser la fusion ou la coopération entre partis partageant des visions similaires peut contribuer à des paysages politiques plus rationalisés et à des systèmes démocratiques plus stables et efficaces. Bien que la présente étude se concentre sur les pays africains, les indices de Noua peuvent être adaptés et appliqués à d'autres régions du monde en ajustant les paramètres du modèle et en intégrant des variables spécifiques à chaque contexte.

\pagebreak

\bibliographystyle{ieeetr}
\renewcommand{\bibname}{References}

\bibliography{references} 

\begin{thebibliography}{10}

\bibitem{ricciuti2004}
R.~Ricciuti, ``Political fragmentation and fiscal outcomes,'' {\em Public
  choice}, vol.~118, no.~3, pp.~365--388, 2004.

\bibitem{pildes2021p}
R.~H. Pildes, ``Political fragmentation and the decline of effective
  government,'' {\em Journal of Democracy}, pp.~22--12, 2021.

\bibitem{RDCCENI}
CENI-RDC,
  ``\href{https://web.archive.org/web/20161206024410/http://ceni.cd/partis_et_regroupements_politiques}{Commission
  Électorale Nationale Indépendante}.''
\newblock Accessed: 2025-02-16.

\bibitem{koffielections}
A.~D. Koffi and A.~K. Agbeto{\'e}zian, ``{\'E}lections en afrique: enjeux
  d{\'e}mocratiques et qu{\^e}te de sens,'' {\em Revue ACAREF - ACADEMIE
  AFRICAINE DE RECHERCHE ET D'ETUDES FRANCOPHONES}, 2021.

\bibitem{tine2017senegal}
A.~Tine, {\em Le S{\'e}n{\'e}gal, sous Senghor et Diouf, une d{\'e}mocratie
  buissonni{\`e}re?: une critique du pluralisme des partis politiques}.
\newblock L'Harmattan S{\'e}n{\'e}gal, 2017.

\bibitem{stoner2013t}
K.~Stoner and M.~McFaul, {\em Transitions to democracy: a comparative
  perspective}.
\newblock JHU Press, 2013.

\bibitem{erdmann2008party}
G.~Erdmann and M.~Basedau, ``Party systems in africa: Problems of categorising
  and explaining party systems,'' {\em Journal of Contemporary African
  Studies}, vol.~26, no.~3, pp.~241--258, 2008.

\bibitem{Osei2006}
A.~Osei, ``La connexion entre les partis et les électeurs en afrique : le cas
  ghanéen,'' {\em Politique africaine}, vol.~104, no.~4, pp.~38--60, 2006.

\bibitem{bratton1997}
M.~Bratton and N.~Van~de Walle, {\em Democratic experiments in Africa: Regime
  transitions in comparative perspective}.
\newblock Cambridge university press, 1997.

\bibitem{Daddieh2014}
C.~K. Daddieh and G.~M. Bob-Milliar, {\em Ghana: the African exemplar of an
  institutionalized two-party system?}, pp.~107--128.
\newblock Springer, 2014.

\bibitem{collier2009}
P.~Collier, {\em Wars, guns and votes: Democracy in dangerous places}.
\newblock Random House, 2009.

\bibitem{bayart2009}
J.-F. Bayart, {\em The State in Africa: the politics of the belly}.
\newblock Wiley, 2009.

\bibitem{Laakso1979}
M.~Laakso and R.~Taagepera, ``Effective number of parties: a measure with
  application to west europe,'' {\em Comparative political studies}, vol.~12,
  no.~1, pp.~3--27, 1979.

\bibitem{taagepera1985rethinking}
R.~Taagepera and B.~Grofman, ``Rethinking duverger's law: predicting the
  effective number of parties in plurality and pr systems--parties minus issues
  equals one,'' {\em European Journal of Political Research}, vol.~13, no.~4,
  pp.~341--352, 1985.

\bibitem{taagepera1993predicting}
R.~Taagepera and M.~S. Shugart, ``Predicting the number of parties: A
  quantitative model of duverger's mechanical effect,'' {\em American Political
  Science Review}, vol.~87, no.~2, pp.~455--464, 1993.

\bibitem{blau2008effective}
A.~Blau, ``The effective number of parties at four scales: Votes, seats,
  legislative power and cabinet power,'' {\em Party politics}, vol.~14, no.~2,
  pp.~167--187, 2008.

\bibitem{caulier2005effective}
J.-F. Caulier and P.~Dumont, ``The effective number of relevant parties: how
  voting power improves laakso-taagepera’s index,'' {\em Munich Personal
  RePEc Archive}, 2005.

\bibitem{grofman2012many}
B.~Grofman and R.~Kline, ``How many political parties are there, really? a new
  measure of the ideologically cognizable number of parties/party groupings,''
  {\em Party Politics}, vol.~18, no.~4, pp.~523--544, 2012.

\bibitem{bhattacharya2006effective}
S.~Bhattacharya and F.~Smarandache, ``Effective number of parties in a
  multi-party democracy under an entropic political equilibrium with floating
  voters,'' {\em Case 1: Shock size 50\% of Y0}, p.~62, 2006.

\bibitem{Golosov2010}
G.~V. Golosov, ``The effective number of parties: A new approach,'' {\em Party
  politics}, vol.~16, no.~2, pp.~171--192, 2010.

\bibitem{borooah2013general}
V.~K. Borooah, ``A general measure of the ‘effective’number of parties in a
  political system,'' in {\em Constitutional economics and public
  institutions}, pp.~146--159, Edward Elgar Publishing, 2013.

\bibitem{taagepera2007}
R.~Taagepera, {\em Predicting party sizes: The logic of simple electoral
  systems}.
\newblock OUP Oxford, 2007.

\bibitem{raymond2016e}
C.~D. Raymond, M.~Huelshoff, and M.~R. Rosenblum, ``Electoral systems, ethnic
  cleavages and experience with democracy,'' {\em International Political
  Science Review}, vol.~37, no.~4, pp.~550--566, 2016.

\bibitem{wolff2005e}
S.~Wolff, ``Electoral systems design and power-sharing regimes,'' {\em
  Powersharing: New Challenges for Divided Societies}, pp.~59--74, 2005.

\bibitem{hazama2003s}
Y.~Hazama, ``Social cleavages and electoral support in turkey: Toward
  convergence?,'' {\em The Developing Economies}, vol.~41, no.~3, pp.~362--387,
  2003.

\bibitem{bogaards2004}
M.~Bogaards, ``Counting parties and identifying dominant party systems in
  africa,'' {\em European journal of political research}, vol.~43, no.~2,
  pp.~173--197, 2004.

\bibitem{kselman2016crowded}
D.~M. Kselman, E.~N. Powell, and J.~A. Tucker, ``Crowded space, fertile ground:
  party entry and the effective number of parties,'' {\em Political Science
  Research and Methods}, vol.~4, no.~2, pp.~317--342, 2016.

\bibitem{golosov2015}
G.~V. Golosov, ``The number of parties and party system nationalization in an
  integrated analytical framework,'' {\em comparative sociology}, vol.~14,
  no.~5, pp.~662--681, 2015.

\bibitem{xhaferaj2014}
A.~Xhaferaj, ``Which parties count?-the effective number of parties in the
  albanian party system,'' {\em Philpapers}, 2014.

\bibitem{li2016seat}
Y.~Li and M.~S. Shugart, ``The seat product model of the effective number of
  parties: A case for applied political science,'' {\em Electoral Studies},
  vol.~41, pp.~23--34, 2016.

\bibitem{ranganathan2004lm}
A.~Ranganathan, ``The levenberg-marquardt algorithm,'' {\em Tutoral on LM
  algorithm}, vol.~11, no.~1, pp.~101--110, 2004.

\bibitem{gcolab}
Google, ``\href{https://colab.google/}{Google Colaboratory}.''
\newblock Accessed: February 27, 2025.

\bibitem{scipy}
SciPy,
  ``\href{https://docs.scipy.org/doc/scipy/reference/generated/scipy.optimize.least_squares.html}{Scipy
  optimize: Least squares Algorithm}.''
\newblock Accessed: February 27, 2025.

\bibitem{WorldBank2025}
W.~B. Group,
  ``\href{https://databank.worldbank.org/source/world-development-indicators}{DataBank,
  World Development Indicators},'' 2025.
\newblock Accessed: 2025-01-30.

\bibitem{ecowas}
ECOWAS, ``\href{https://ecowas.int/}{Economic Community of West African
  States}.''
\newblock Accessed: 2025-02-16.

\bibitem{eccas}
ECCAS, ``\href{https://ceeac-eccas.org/}{Economic Community of Central African
  States }.''
\newblock Accessed: 2025-02-16.

\bibitem{allianceofsahelstates}
AES,
  ``\href{https://en.wikipedia.org/wiki/Alliance_of_Sahel_States}{Confederation
  of Sahel States}.''
\newblock Accessed: 2025-02-16.

\end{thebibliography}

\end{document}